\documentclass[12pt]{article}

\usepackage[latin1]{inputenc}
\usepackage{amsmath}
\usepackage{amsfonts}
\usepackage{amssymb,MnSymbol,slashed}
\usepackage{amsxtra}
\usepackage[margin=3cm]{geometry}
\usepackage{color}
\usepackage{hyperref}
\hypersetup{linktocpage=true}
\usepackage{graphicx}
\usepackage{placeins}
\usepackage{upgreek} 
\usepackage{mathrsfs}
\usepackage{accents}
\usepackage{multirow} 
\usepackage{tikz}
\usepackage{braket}

\usepackage{cite}

\numberwithin{equation}{section}
\newcommand{\beq}{\begin{equation}}
\newcommand{\eeq}{\end{equation}}
\def\be {\begin{equation}}
\def\ee {\end{equation}}
\def\bs#1\es{\begin{split}#1\end{split}}
\def\ba#1\ea{\begin{align}#1\end{align}}
\def\baed#1\eaed{\begin{aligned}#1\end{aligned}}
\def\bged#1\eged{\begin{gathered}#1\end{gathered}}
\def\bea{\begin{eqnarray}}
\def\eea{\end{eqnarray}}

\newcommand{\nn}{\nonumber}

\def\d{\delta}

\def\e{\epsilon}

\def\G{\Gamma}

\def\m{\mu}
\def\n{\nu}
\def\o{\omega}

\newcommand{\cT}{\mathcal{T}}

\newcommand{\cK}{\mathcal{K}}
\newcommand{\cW}{\mathcal{W}}
\newcommand{\cG}{\mathcal{G}}

\newcommand{\cR}{\mathcal{R}}

\def\cO{{{\mathcal O}}}
\def\cM{\mathcal{M}} 
\def\cN{\mathcal{N}}
\def\cV{\mathcal{V}}

\def\Tr{\text{Tr}}

\def\pa{\partial}

\def\fr{\frac}

\def\we{\wedge}

\def\tbzero{{\text{\tiny{(0)}}}}
\def\tbone{{\text{\tiny{(1)}}}}

\newcommand{\wh}[1]{ {\hat{#1}}{} }

\let\foo\bar 
\renewcommand{\bar}[1]{ {\foo{  #1} }{} }

\newlength{\dhatheight}

\begin{document}

\baselineskip=16pt
\setlength{\parskip}{6pt}

\begin{titlepage}
\begin{flushright}
\parbox[t]{1.4in}{
\flushright
IPMU16-0115
YITP-SB-16-32}
\end{flushright}

\begin{center}

\vspace*{1.5cm}

{\Large \bf  The Euler characteristic correction\\[2mm] to the K\"{a}hler potential---revisited} 

\vskip 1.5cm

\renewcommand{\thefootnote}{}

\begin{center}
 \normalsize 
 Federico Bonetti\textsuperscript{1 \!\!\!\dagger}\footnote{\textsuperscript{\dagger}bonetti@insti.physics.sunysb.edu}, Matthias Weissenbacher\textsuperscript{2 \!\!\!\ddagger}\footnote{\textsuperscript{\ddagger}matthias.weissenbacher@ipmu.jp}
\end{center}
\vskip 0.5cm
\textsuperscript{1}C.N. Yang Institute for Theoretical Physics, SUNY Stony Brook,\\
Stony Brook, New York 11794, U.S.A.

\textsuperscript{2}Kavli Institute for the Physics and Mathematics of the Universe, University of Tokyo,
Kashiwa-no-ha 5-1-5, 277-8583, Japan
\end{center}

\vskip 1.5cm
\renewcommand{\thefootnote}{\arabic{footnote}}

\begin{center} {\bf ABSTRACT } \end{center}
We confirm the    leading  $\alpha'^3$ correction to the 4d, $\mathcal N = 1$ K\"{a}hler potential
of type IIB orientifold compactifications,  proportional to the 
Euler characteristic of the Calabi-Yau threefold (BBHL correction).
We present the explicit solution for the $\alpha'^3$-modified internal background metric
in terms of the non-harmonic part of the third Chern form of the leading order Calabi-Yau manifold. The corrected internal manifold
 is almost Calabi-Yau and admits an
$SU(3)$ structure with non-vanishing torsion.
We also find that the full ten-dimensional  Einstein frame background metric is multiplied 
by a non-trivial Weyl factor.
Performing a Kaluza-Klein reduction on the modified background we derive the $\alpha'^3$-corrected 
 kinetic terms for the dilaton and the K\"{a}hler deformations of the internal Calabi-Yau threefold for arbitrary $h^{1,1}$. 
We analyze these kinetic terms in the 4d, $\mathcal N = 2$ un-orientifolded  theory,
confirming the expected correction to the K\"ahler moduli space prepotential,
as well as in
the 4d, $\mathcal N = 1$ orientifolded theory,
thus determining the corrections to the K\"ahler potential
and K\"ahler coordinates.

\end{titlepage}

\newpage
\noindent\rule{\textwidth}{.1pt}		
\tableofcontents
\vspace{20pt}
\noindent\rule{\textwidth}{.1pt}

\setcounter{page}{1}
\setlength{\parskip}{9pt}

\section{Introduction}

Supersymmetric flux compactifications of type IIB superstring theory
constitute a vast and rich arena for the study of the implications 
of string dynamics on 4d physics. Comprehensive reviews
on the subject include \cite{Grana:2005jc, Douglas:2006es,
Blumenhagen:2006ci, Denef:2007pq , Samtleben:2008pe}.
A common feature of several flux compactification scenarios
is the crucial role played by $\alpha'$ and/or $g_s$
corrections to the leading-order effective action.
One of the most prominent perturbative corrections
to the 4d low-energy effective action is the $\alpha'^3$
correction to the K\"ahler potential proportional to the Euler
characteristic of the Calabi-Yau threefold used in the compactification.
This correction was first computed in a paper
by Becker, Becker, Haack, and Louis (BBHL) \cite{Becker:2002nn}.
It plays in essential role 
in the Large Volume Scenario for type IIB compactifications \cite{Balasubramanian:2005zx,
Conlon:2005ki, Cicoli:2008va}.

The 10d origin of the Euler characteristic correction to the 4d K\"ahler
potential resides in the leading $\alpha'$ corrections to the 
bulk type IIB supergravity action. This is an eight-derivative coupling
built with four Riemann tensors,    accompanied 
by several additional couplings involving other 
type IIB fields as a consequence of $SL(2,\mathbb Z)$ symmetry.
For the derivation and discussion of these
couplings see e.g.~\cite{Grisaru:1986dk, Grisaru:1986kw, Schwarz:1982jn,
Gross:1986iv, Gross:1986mw, Sakai:1986bi,   Abe:1987ud,
Kehagias:1997cq, Kehagias:1997jg, Minasian:2015bxa}.
The authors of \cite{Becker:2002nn}
infer the corresponding correction to the 4d K\"ahler  potential
in a somewhat indirect way. 
Exploiting mirror symmetry, the starting point of their analysis 
are the results of \cite{Ferrara:1989ik,Candelas:1990rm} about the corrected 
metric of the hypermultiplet moduli space of a Calabi-Yau type II compactification
to four dimensions with $\cN = 2$ supersymmetry.
Making use of the results of \cite{Bohm:1999uk} these corrections are then
reformulated in terms of 4d field variables whose 10d origin
is manifest, in such a way that the orientifold projection to $\cN = 1$
supersymmetry can be performed.

The aim of this paper is to provide a direct derivation of the Euler characteristic
correction to the 4d K\"ahler potential by means of a completely explicit
Kaluza-Klein reduction of the relevant   $\alpha'^3$ couplings
in the 10d bulk action.   
Not only is this approach more transparent, but it also presents 
other advantages. Firstly, we do not have to make any assumption 
about the superpotential or the scalar potential of the full
4d, $\cN = 1$ low-energy effective action, since our conclusions are 
entirely drawn from the examination of the kinetic terms
for the dilaton and the K\"ahler moduli of the Calabi-Yau threefold for arbitrary $h^{1,1}$.
Secondly, our approach
 allows us to derive both the 
correction to the 4d K\"ahler potential and the 
correction to the K\"ahler coordinates as a function of the
K\"ahler moduli.
Finally, in the process of the derivation one necessarily has 
to  analyze the
background solution and  show how it gets modified
by the $\alpha'$ corrections under examination.
In particular, we are able to provide an explicit solution for the
$\alpha'$-corrected internal metric in terms of the 
non-harmonic part of the third Chern form $c_3$
of the leading order Calabi-Yau threefold background.
The corrected metric is K\"ahler but not Ricci-flat.
As a result it no longer has $SU(3)$ holonomy, but rather an
$SU(3)$ structure with non-vanishing torsion.
The corrected geometry fits in the framework
of almost Calabi-Yau manifolds \cite{Joyce:2001xt}.
Similar results were derived in \cite{Grimm:2014xva} in the case
of M-theory compactifications on a Calabi-Yau fourfold.
Additionally, we find that the 10d background metric
is corrected by an overall Weyl factor at order $\alpha'^3$,
in analogy with the findings of the three-dimensional 
M-theory analysis of \cite{Grimm:2014xva, Grimm:2014efa, Grimm:2015mua, Grimm:2013gma, Grimm:2013bha}.
Throughout this paper, we let higher derivatives
act in the internal space, working at two-derivative level in the external
directions. Recently, $\alpha'^3$ external four-derivative
couplings between K\"ahler moduli and gravity were derived in 
\cite{Weissenbacher:2016gey}
for the same class of type IIB orientifold setups considered in this work,
with potential applications to    K\"ahler moduli inflation studies.

The main results of our paper are summarized in equations \eqref{final1}
to \eqref{final5}.
We  reproduce the Euler characteristic correction to the K\"ahler
potential originally found by BBHL \cite{Becker:2002nn}.
We also compute the leading $\alpha'^3$ corrections to the K\"ahler coordinates
and the 4d axio-dilaton
in terms of the K\"ahler moduli,
and find that they are vanishing. 
Finally, we exclude the possibility that
the effect of the correction to the K\"ahler potential
can be undone by a K\"ahler coordinate redefinition,
and 
we reformulate our findings in the formalism of
linear multiplets  \cite{Grimm:2004uq, Grimm:2004ua, Grimm:2005fa}
in order to elucidate the physical relevance of the correction.
Let us remark that our analysis does not take into account explicitly
  localized sources of the $\cN =1$ setup, such as seven-branes and orientifold planes.

This paper is organized as follows. Section \ref{4dLagr} 
is devoted to the computation of the relevant $\alpha'$ corrections
to the 4d kinetic terms of the dilaton and K\"ahler moduli.
In particular, in section  \ref{higher_deriv_section} 
we present the relevant
higher-derivative corrections to the 10d action of type IIB supergravity, 
in section \ref{background_section} we discuss
the background solutions of interest, while section \ref{redres}
is devoted to the dimensional reduction
to four dimensions. 
The results of the computation are 
then analyzed in section \ref{N2section} in the context of the 4d, $\cN = 2$ effective theory
obtained in the absence of orientifold planes and D-branes.
We show that our findings are compatible with the expected
perturbative correction to the prepotential for the K\"ahler moduli space metric.
We then proceed   
 in section \ref{Kaehler_section} to analyze the 4d, $\cN =1$
setup obtained after implementation of an orientifold projection. 
We verify that the $\alpha'$-corrected kinetic terms
of the dilaton and K\"ahler moduli can be written in terms of a corrected
K\"ahler potential. We determine the latter, reproducing the 
correction of BBHL 
\cite{Becker:2002nn},  and we  identify the $\alpha'$-corrected form of the
K\"ahler coordinates. Section \ref{conclusions_section} summarizes our conclusions.
Our conventions, together with useful identities and some technical
material, can be found in the appendices.

 
\section{Four-dimensional $\alpha'^3$ Lagrangian}\label{4dLagr}

 This section discusses the dimensional reduction of
IIB supergravity action including a suitable class of  eight-derivative corrections, on a Calabi-Yau threefold to four dimensions.
In particular, we restrict our analysis to 
  purely gravitational terms  and dilaton terms. 
   We   fluctuate the background metric by K\"{a}hler deformations and focus on couplings that carry two external spacetime derivatives and are at most quadratic in the infinitesimal 
   K\"{a}hler deformations.
We first review the relevant eight-derivative $\alpha'^3$
corrections to    10d
type IIB supergravity and the supersymmetric background.

\subsection{Type IIB higher-derivative action}\label{higher_deriv_section}

The starting point of our analysis consists of  the type IIB supergravity
action  including the  leading order $\alpha'^3$ eight-derivative
correction  built with four Riemann tensors
\cite{Grisaru:1986dk, Grisaru:1986kw, Schwarz:1982jn,
Gross:1986iv, Gross:1986mw, Sakai:1986bi,   Abe:1987ud,
Kehagias:1997cq, Kehagias:1997jg, Minasian:2015bxa}.
For our purposes this gravitation coupling has to be supplemented with suitable
dilaton couplings discussed below.

 In order to set up our notation, let us first  record the gravitational
and axio-dilaton terms 
 in the two-derivative type IIB supergravity in the Einstein frame,
\beq \label{twoderivative}
S^{(2)}_{R, \tau} = \frac{1}{2\kappa_{10}^2} \int \left( 
R  -  \frac{1}{2\tau_2^2} \nabla_M \tau \nabla^M \overline \tau
\right) *_{10}1  \ ,
\eeq
where $2\kappa_{10}^2 = (2\pi)^7 \alpha'^4$, $M = 0,\ldots ,9$
is a 10d world index, and the axio-dilaton
is defined as
\beq
\tau = \tau_1 + i \, \tau_2  = C_0 + i \, e^{- \wh\phi} \ .
\eeq
The action \eqref{twoderivative} is   invariant under the  
 $SL(2,\mathbb Z)$ symmetry of type IIB superstring theory,
under which the Einstein frame metric is invariant and the 
axio-dilaton transforms as
\beq
\tau ' = \frac{a \, \tau + b}{c \, \tau + d} \ , \qquad
\begin{pmatrix}
a & b \\
c & d
\end{pmatrix} \in SL(2,\mathbb Z) \ .
\eeq
Next, let us consider the $\alpha'^3$ correction constructed with
four Riemann tensors. It takes the form 
\beq \label{R4actionpartial}
S^{(8)}_{R^4} = \frac{1}{2\kappa_{10}^2} \cdot \frac{\alpha'^3}{3 \cdot 2^{12}}
\int    f_0(\tau , \overline \tau) \left( t_8 t_8 + \tfrac{1}{8} \epsilon_{10}
\epsilon_{10} \right) R^4 \, *_{10}1 \ ,
\eeq
where the explicit tensor contractions are given by
\ba\label{deft8t8R4temporary}
 \e_{10}  \e_{10}   R^4 &=  \epsilon^{R_1 R_2   M_1\ldots M_{8} }  \epsilon_{R_1 R_2 N_1 \ldots N_{8}}   R^{N_1 N_2}{}_{M_1 M_2}   R^{N_3 N_4}{}_{M_3 M_4}    R^{N_5 N_6}{}_{M_5 M_6}   R^{N_7 N_8}{}_{M_7 M_8} \ , \nonumber \\[.1cm]
 t_8  t_8   R^4 & =   t_{8}^{  M_1 \dots M_8}   t_{8 \, N_1  \dots N_8}       R^{N_1 N_2}{}_{M_1 M_2}   R^{N_3 N_4}{}_{M_3 M_4}    R^{N_5 N_6}{}_{M_5 M_6}   R^{N_7 N_8}{}_{M_7 M_8} \ ,
\ea
and the tensor $t_8$ is defined in terms of the metric in the standard way
 \cite{Schwarz:1982jn}.
We have also introduced the function $f_0(\tau , \overline \tau)$, which is the non-holomorphic, 
$SL(2,\mathbb Z)$-invariant function defined as
\beq
f_0(\tau , \overline \tau) =
 \sum_{(m,n) \neq(0,0)} \frac{\tau_2^{3/2}}{|m + n \, \tau|^3} \ .
\eeq
It is useful to note that in the large $\tau_2$ limit, which corresponds to the small
string coupling limit, we have
\beq \label{f0expansion}
f_0(\tau ,\overline \tau)  = 2 \zeta(3) \, \tau_2^{3/2} + \frac{2\pi^2}{3} \tau_2^{-1/2}
+ \cO(e^{-2\pi \tau_2}) \ .
\eeq 

Our analysis requires the consideration of additional
higher-derivative terms involving gradients of the dilaton.
Such terms can be obtained  following the 
 approach of  \cite{Gross:1986mw, Kehagias:1997cq ,Kehagias:1997jg }, 
 by replacing each occurrence of the Riemann tensor
in \eqref{deft8t8R4temporary}   according to
\beq \label{Rreplacement}
{R_{MN}}^{PQ} \rightarrow    R_{MN}{}^{PQ}
+ \tilde c_1 g_{[M} {}^{[P} \nabla_{N]} \nabla^{Q]} \wh \phi 
+ \tilde c_2 g_{[M} {}^{[P} \nabla_{N]} \wh \phi  \nabla^{Q]} \wh \phi 
+ \tilde c_3 g_{[M}{}^{[P} g_{N]}{}^{Q]} \nabla_K \phi \nabla^K \wh \phi
 \ ,
\eeq
where antisymmetrizations are performed with weight one, and $\tilde c_{1}$,
$\tilde c_{2}$,
$\tilde c_{3}$ are real numerical coefficients that we leave unfixed for now.\footnote{The analysis
of the 4-point tree-level dilaton scattering amplitude
gives the value $\tilde c_1 = -1$ (see \cite{Gross:1986mw,Kehagias:1997cq ,Kehagias:1997jg} and also \cite{Policastro:2006vt , Policastro:2008hg}),
while $\tilde c_2$, $\tilde c_3$ cannot be fixed in this way.
In what follows, however, we proceed treating all $\tilde c$ coefficients
on the same footing.}
In order to achieve $SL(2,\mathbb Z)$-invariance several other  terms
have to be added  to the action \cite{Policastro:2006vt , Policastro:2008hg},
but they are not relevant for our discussion.

In summary, the total action utilized in the analysis of the 
following sections is obtained by summing the two-derivative
terms in \eqref{twoderivative} with the terms generated by the
replacement rule  \eqref{Rreplacement} in the 
terms given in  \eqref{deft8t8R4temporary}.
We  only keep terms which can correct the kinetic terms of the resulting 4d dilaton and discarding higher derivative terms thereof, as well as terms for the axion $C_0$. We also
retain only the leading term in the large $\tau_2$ expansion
of $f_0(\tau, \overline \tau)$, see \eqref{f0expansion}.
We thus obtain
\begin{align} \label{R4actiontot}
S  = \frac{1}{2\kappa_{10}^2} \int \bigg[
& R - \frac 12 \nabla_M \phi \nabla^M \phi
+ \frac{\zeta(3) \alpha'^3}{3 \cdot 2^{11}} \, e^{-\frac 32 \phi}
\left( t_8 t_8 + \frac 18 \epsilon_{10} \epsilon_{10} \right) R^4  \\
& + \frac{\zeta(3) \alpha'^3}{3 \cdot 2^{11}}
\, e^{- \frac 32 \phi}
\bigg(  
\tilde c_1 \nabla^M  \nabla_M \wh \phi \, f_1
+ \tilde c_2  \, \nabla_M \wh \phi \, \nabla_N \wh \phi \, f_2^{MN}
+ \tilde c_3  \, \nabla^M  \wh \phi \, \nabla_M \wh \phi \, f_3
\bigg) \bigg] *_{10} 1  \ , \nn
\end{align}
where the quantities $f_1$, $f_2^{MN}$, $f_3$
are homogeneous polynomials of degree three in the 
components of the Riemann tensor and are given explicitly in appendix \ref{app action}.
 For notational convenience let us introduce the dimensionful
 constant
 \beq \label{alpha_def}
\alpha = \frac{\zeta(3) \alpha'^3}{3 \cdot 2^{11}}  \ ,
 \eeq
which plays the role of the  small expansion parameter relevant for our problem.
Note that we adopt conventions in which the 10d metric components and  the dilaton are dimensionless,
while coordinates $x^M$ have dimension of length.

\subsection{Supersymmetric background}\label{background_section}

In this section we study how the supersymmetric Calabi-Yau background 
solutions of type IIB supergravity are modified in the presence of the higher-derivative
terms in \eqref{R4actiontot} and we present the relevant dilaton and K\"ahler 
fluctuations around the corrected background. Let us note that we cannot
analyze directly the supersymmetry properties of the background, since
the type IIB supersymmetry variations are not completely known at the required
order in $\alpha'$, but we can 
 give necessary conditions for the $\alpha'$ modification of the background   by solving the equations of motion.

\subsubsection{Corrections to  the background} \label{background_without_fluctuations}

Our problem fits into the framework of  
supersymmetric flux compactifications of type IIB superstring
on a Calabi-Yau
threefold $Y_3$. Let us first recall some facts about 
these setups 
neglecting higher-derivative corrections to the 10d supergravity
action \cite{Giddings:2001yu}.
For compactifications to flat four-dimensional spacetime  
the background metric has the form
\ba\label{metricbackg}
ds^2_{10} =e^{2A} \,  \eta_{\m \n} dx^\m dx^\n 
+ e^{-2A} \, g^\tbzero_{ab} d y ^a d y^b   \;\; ,
\ea
where
 $\m, \nu = 0,1,2,3$  are 4d external spacetime world indices, $x^\mu$
 are Cartesian coordinates,
$\eta_{\mu\nu}$ is the Minkowski metric,
$a,b = 1, \dots 6$ are real internal world indices
associated to the coordinates $y^a$,
and $g^\tbzero_{ab}$ denotes the Ricci-flat metric on $Y_3$.
The warp factor $A$ only depends on the internal coordinates
and also determines the background $F_5$ flux via
\beq
F_5 =(1 + *_{10})  \, d e^{4A} \wedge dx^0 \wedge dx^1 \wedge dx^3 \wedge dx^4 \ .
\eeq
The $F_5$ flux has to obey the Bianchi identity
\beq \label{F5Bianchi}
dF_5 = H_3 \wedge F_3 + \rho_6 \ ,
\eeq
where $H_3 = dB_2$ and $F_3  = dC_2 - C_0 dB_2$
are the usual NSNS and RR three-form
field strengths and the six-form $\rho_6$ encodes the D3-brane charge density
associated with possible localized sources.
Integrating \eqref{F5Bianchi} on the internal manifold yields the D3-brane
tadpole cancellation condition
\beq \label{tadpole_cancellation}
\frac{1}{2 \kappa_{10}^2 T_3}\int_{Y_3} H_3 \wedge F_3 + Q_{\rm D3} = 0 \ ,
\eeq
where $T_3 = (2\pi)^{-3} (\alpha')^{-2}$ is the D3-brane tension and
 $Q_{\rm D3}$ is the total D3-brane charge, proportional to the integral of 
 $\rho_6$.\footnote{Recall that $Q_{\rm D3}$ generically receives contributions
not only from $D3$-branes and $O3$-planes, but also from higher-dimensional
defects with flux- and/or geometry-induced $D3$-brane charge,
such as $D7$-branes and $O7$-planes\cite{Giddings:2001yu}.}
 If all local sources are removed,
all fluxes have to vanish and the warp factor is necessarily trivial.

The inclusion of higher-derivative corrections to the type IIB bulk action
induces modifications in the previous picture. 
In what follows we analyze the 10d equations of motion in presence
of higher-derivative corrections of order $\alpha \propto \alpha'^3$
but without introducing any local source in the problem. 
Even though this setup yields an effective 4d, $\cN = 2$
low-energy effective action, we argue that the following analysis 
is sufficient for the purpose of studying the Euler characteristic
correction to the K\"ahler potential in the   4d, $\cN = 1$ theory after performing the orientifold truncation in section \ref{Kaehler_section}.

We adopt the following $\alpha'$-corrected Ansatz for the 10d background metric,
\beq \label{full_background_metric_Ansatz}
ds^2_{10} = e^{\Phi} \left[
e^{2A} \, \eta_{\mu\nu} dx^\mu dx^\nu
+ e^{-2A} \, g_{ab} d y ^a d y^b
\right] \ ,
\eeq
where
\bea
\Phi & =&  \alpha \, \Phi^\tbone + \cO(\alpha^2) \ , \nn \\
A & =&  \alpha \, A^\tbone + \cO(\alpha^2) \ , \nn \\
g_{ab} & =& g^\tbzero_{ab}  
+ \alpha \, g_{ab}^\tbone + \cO(\alpha^2) \ .
\eea
The quantity $\Phi$ is an overall 10d Weyl rescaling of the metric,
and it has been introduced in analogy with the analysis of \cite{Grimm:2014xva}.
This parametrization of the prefactors multiplying the internal and external metric
is   general and is useful for the following discussion.
The zeroth-order metric is the Calabi-Yau Ricci-flat metric.
Accordingly, $\Phi$ and $A$ have no 
$\cO(\alpha^0)$ term.
All quantities depend exclusively on the internal coordinates in order not to
break Poincar\'e invariance in the external directions.
The corrections to the dilaton profile are parametrized as
\beq
\wh \phi = \phi_0  + \alpha \, \phi^\tbone + \cO(\alpha^2) \ ,
\eeq
where $\phi_0$ is the constant uncorrected dilaton VEV.

The 10d Einstein equation at order $\alpha$ can be written in the form
\beq
0 = R^\tbone_{MN} - \tfrac 12  (R_{PQ}^\tbone \, g^{\tbzero PQ})  \,
  g^\tbzero_{MN} + \cT_{MN} \ .
\eeq
The first two terms capture the contribution coming from the 
  two-derivative part of the Einstein equation
evaluated on the $\alpha$-corrected Ansatz  \eqref{full_background_metric_Ansatz}.
The symbol    $R^\tbone_{MN}$ is used to denote
the $\cO(\alpha)$ part of the 10d Ricci tensor
computed using the metric   \eqref{full_background_metric_Ansatz},
while $g^\tbzero_{MN}$ is used 
for the $\cO(\alpha^0)$ part of  \eqref{full_background_metric_Ansatz}.
The quantity $\cT_{MN}$ encodes all the contributions coming from the
higher-derivative part of the Einstein equation, derived from \eqref{R4actiontot},
upon evaluation on the $\cO(\alpha^0)$ part of the 
metric Ansatz \eqref{full_background_metric_Ansatz}.
We find
\beq
\cT_{\mu\nu} = 0  \ , \qquad \cT_{\mu a} = 0 \ , \qquad
\cT_{ab} =768 \, \alpha \, e^{-\frac 32 \phi_0 }\, J^\tbzero{}_a{}^{c} J^\tbzero{}_b{}^{d} \,
  \nabla^\tbzero_{c} \nabla^\tbzero_{d} Q \ , 
\eeq
where $\nabla^\tbzero$, $J^\tbzero$ denote the Levi-Civita connection and
 complex structure associated to the
zeroth-order metric, respectively,
and the quantity $Q$ is the six-dimensional Euler density,
\beq
Q =\tfrac {1}{12} \left( R^\tbzero_{a_1 a_2}{}^{a_3 a_4} 
R^\tbzero_{a_3 a_4}{}^{a_5 a_6} 
R^\tbzero_{a_5 a_6}{}^{a_1 a_2}
- 2 R^\tbzero_{a_1}{}^{a_2}{}_{b_1}{}^{b_2} 
R^\tbzero_{a_2}{}^{a_3} {}_{b_2} {}^{b_3}
R^\tbzero_{a_3} {}^{a_1} {}_{b_3} {}^{b_1}  \right) \ .
\eeq
This object satisfies
\beq \label{Q_integral}
Q = (2\pi)^3  \, *^\tbzero_6 c^\tbzero_3 \ , \qquad
\int Q *^\tbzero_6 1 = (2\pi)^3 \chi \ ,
\eeq
where $\chi$ is the Euler characteristic of the internal space,
$*^\tbzero_6$ is the Hodge star operator with respect to the 
zeroth-order metric, and $c_3$ is the third Chern form
built from $g^\tbzero_{ab}$, defined in appendix \ref{Conv_appendix}.

It is convenient to use holomorphic and antiholomorphic indices
$m = 1,2,3$, $\bar m = \bar 1, \bar 2, \bar 3$ associated to the 
zeroth-order complex structure, in such a way that $J^\tbzero{}_m{}^n = + i \delta_{m}{}^n$. 
One may then check that 
all components of the order-$\alpha$ Einstein's equation are solved by imposing
\begin{align} \label{Weyl_factor}
\Phi^\tbone = -192 \,  e^{-\frac 32 \phi_0} \, Q \  , \quad
R^\tbone_{m \bar n} = - 1536 \,  e^{-\frac 32 \phi_0} \, \nabla^\tbzero_m
\nabla^\tbzero_{ \bar n} Q 
\ , \quad 
R^\tbone_{mn} = 0 \ , \quad
R^\tbone_{\bar m \bar n} = 0 \ , \quad 
  A^\tbone =0 \ ,
\end{align}
where $R^\tbone$ denotes the linearized Ricci tensor of the metric
correction $g^\tbone$.
Let us also point out that
we can exhibit an explicit expression
for $g^\tbone$.
To this end we start recalling that, as a consequence of the
$\partial \bar \partial$-lemma, the $(3,3)$-form $c_3$ can be decomposed as
\beq \label{deldelbarconsequence}
c_3 = H c_3 + i \partial \bar \partial F \ ,
\eeq
where $H c_3$ denotes the harmonic part of $c_3$ with respect to the zeroth-order
metric and $F$ is a suitable co-closed $(2,2)$-form.\footnote{
The existence and co-closure of $F$ can be seen by the following argument,
similar to an argument in \cite{Nemeschansky:1986yx}. 
Let us first apply the Hodge decomposition theorem to the 6-form $c_3$
and write $c_3 = Hc_3 + db_5$ for a suitable globally defined 5-form
$b_5$. Since the form $db_5$ is $d$-exact, $\partial$-closed, and $\bar \partial$-closed,
the $\partial \bar \partial$-lemma ensures that it is $\partial \bar \partial$-exact,
so that $d\beta_5 = \partial \bar \partial f_4$ for a globally defined 4-form $f_4$.
We can now apply the Hodge decomposition theorem to $f_4$ and write
$f_4 = Hf_4 + dg_3 + \delta g_5$, where $\delta$ denotes the codifferential
and $g_3$, $g_5$ are a globally defined 3- and 5-form, respectively.
We now note that $H f_4$ is harmonic on a K\"ahler manifold and hence $\bar \partial$-closed, and that $\partial \bar \partial d = 0$. It follows that we can write
$c_3 = Hc_3 + \partial \bar \partial \delta g_5$.
We can thus set $F= -i \delta g_5$, which is co-closed because it is co-exact.
}
We can express the Euler density $Q$ in terms of $F$ as
\beq \label{QintermsofF}
(2\pi)^{-3} \, Q =  *^\tbzero_6 H c_3 + \,\tfrac 12   \, \Delta^\tbzero  *^\tbzero_6 (J^\tbzero \wedge F) \ ,
\eeq
where $\Delta^\tbzero = 2 g^{\tbzero m \bar n} \nabla^\tbzero_m \nabla^\tbzero_{\bar n}$ denotes the scalar Laplacian
of the zeroth-order metric, and we exploited the co-closure of $F$.
Let us also stress that the first term in \eqref{QintermsofF}
is constant on the threefold as a result of the harmonic projection.
Utilizing \eqref{QintermsofF} one can observe that 
the equations for $R^\tbone$ in \eqref{Weyl_factor} 
can be solved by setting
\beq \label{gtbone_solution}
g^\tbone_{m n} = 0  \ , \quad
g^\tbone_{\bar m \bar n} = 0 \ , \quad 
g^\tbone_{m \bar n} = -1536 (2\pi)^3 \, e^{-\frac 32 \phi_0} \;  \nabla^\tbzero_m \nabla^\tbzero_{\bar n}
*^\tbzero_6 (J^\tbzero \wedge F)\ .
\eeq
As we can see, the corrected metric $g^\tbzero_{m\bar n}
 + \alpha g^\tbone_{m \bar n}$
is still K\"ahler and belongs to the same K\"ahler class as $g^\tbzero_{m \bar n}$.

The dilaton also receives a correction sourced by
the higher-derivative terms in \eqref{R4actiontot}.
The relevant terms in the dilaton equation of motion  yield
the relation 
\beq \label{dilatonEOM}
0 = g^\tbzero{}^{ \bar n m } \nabla^\tbzero_m \nabla^\tbzero_{\bar n}  \phi^\tbone
- 1152   \, e^{-\frac 32 \phi_0} \, \tilde c_1  \, 
 g^\tbzero{}^{\bar n m} \nabla^\tbzero_m \nabla^\tbzero_{\bar n} Q\ ,
 \eeq
and is solved by
\beq \label{dilaton_correction}
 \phi^\tbone =1152 \,  \tilde c_1  \, e^{-\frac 32 \phi_0}  Q \ .
\eeq
An integration constant has been set to zero because, 
up to $\cO(\alpha^2)$  terms,
it can always be 
reabsorbed in $\phi_0$.
Our findings about the modification of the background
dilaton profile and internal metric
are in line with previous work \cite{Gross:1986iv,
Freeman:1986br,Freeman:1986zh,Candelas:1986tz,Antoniadis:1997eg}.

In closing this section let us remark that a full determination of
 $F_1$, $G_3$, $F_5$ at order $\alpha$
would require a detailed knowledge of 
order-$\alpha$ corrections to their equations of motion,
which are related to the $SL(2,\mathbb Z)$
completion
of the higher-derivative terms recorded in \eqref{R4actiontot}.
This investigation is beyond the scope of the present work.
Since we  focus on couplings involving the metric and the dilaton,
however, we only need the expressions \eqref{Weyl_factor}
and \eqref{dilaton_correction} for the order-$\alpha$ Weyl rescaling factor and dilaton
correction.

\subsubsection{$SU(3)$ structure on the internal manifold}

In this section we show how the $SU(3)$ holonomy of the zeroth-order Calabi-Yau
internal metric is deformed to a specific $SU(3)$ structure 
 of the  $\alpha'^3$-modified background.
Our discussion applies to $g^\tbzero_{ab} + \alpha \, g^\tbone_{ab}$
and does not involve the  overall Weyl rescaling factor of the 10d metric.

The K\"ahler form and the $(3,0)$-form of the uncorrected Calabi-Yau
metric are subject to $\alpha'$ corrections but remain nonetheless
globally defined on the internal space. As a result, 
the structure group of the internal manifold is reduced from $SO(6)$
to $SU(3)$ even after taking corrections  into account. This guarantees 
 the existence of local $SU(3)$ coframes,
consisting of triplets $e^I$, $I =1,2,3$, of complex 
one-forms together with their complex conjugates 
$\bar e^{\bar I}$,
with the property that $\{ e^I  ,  \bar e^{\bar I} \}$
is a local basis of $T^*Y_3 \otimes \mathbb C$ and
on overlapping patches the transition functions for $e^I$
take values in $SU(3)$. We may express locally $J$ and $\Omega$, as well as the metric,
in terms of the $SU(3)$ coframe as
\beq
J = i \, \delta_{I \bar J} \, e^I \wedge \bar e^{\bar J} \ , \qquad
\Omega = \tfrac {1}{3!} \, \epsilon_{IJK} \, e^I \wedge e^J \wedge e^K \ , \qquad
ds^2 = 2 \, \delta_{I \bar J} \, e^I \, \bar e^{\bar J} \ ,
\eeq 
where $\epsilon_{123} = 1$.
The coframe can be written as the sum of an uncorrected contribution
and a correction,
\beq
e^I = e^\tbzero{}^I + \alpha \, e^\tbone{}^I  \ .
\eeq
Since the uncorrected geometry is K\"ahler, we can
adopt complex coordinates and choose the coframe $e^\tbzero{}^I$
in such a way that 
its
  only non-zero components are $e^\tbzero{}^I{}_m$,
$\bar e^\tbzero{}^{\bar I} {}_{\bar n}$.

The relevant correction to the $SU(3)$ coframe takes the form
\beq
e^\tbone{}^I{} =- 768 (2\pi)^3 \, e^{-\frac 32 \phi_0} \,  e^\tbzero{}^I{}_n \, \nabla^\tbzero_m 
\nabla^\tbzero{}^n   *^\tbzero_6 (J^\tbzero \wedge F)  \; dz^m \ .
\eeq
It is indeed straightforward to check that the associated correction to the metric
reproduces $g^\tbone$ in \eqref{gtbone_solution}. We also have
\beq
J = J^\tbzero + \alpha \, J^\tbone \ , \qquad 
J^\tbone = - 1536 i \, (2\pi)^3 e^{-\frac 32 \phi_0} \, \partial^\tbzero \bar \partial^\tbzero  
 *^\tbzero_6 (J^\tbzero \wedge F)  \ ,
\eeq
and therefore
\beq
d J^\tbone = 0 \ .
\eeq
The correction to the $(3,0)$ form reads
\beq
\Omega = \Omega^\tbzero + \alpha \, \Omega^\tbone \ , \qquad
\Omega^\tbone = \left[  - 384 \, (2\pi)^3 e^{-\frac 32 \phi_0}  \, 
\Delta^\tbzero  *^\tbzero_6 (J^\tbzero \wedge F) 
\right] \Omega^\tbzero  \ ,
\eeq
which implies
\beq
d\Omega^\tbone =  - 768 \, (2\pi)^3 e^{-\frac 32 \phi_0} \; dQ \wedge \Omega^\tbzero  \ .
\eeq
We can summarize our conclusions in the language of $SU(3)$ torsion classes,
reviewed for instance in \cite{House:2005yc}. 
The most general $SU(3)$ structure can be described by
\begin{align}
dJ & = - \tfrac 32 {\rm Im}(\cW_1 \, \overline \Omega)
+ \cW_4 \wedge J  + \cW_3 \ , \nn \\
d\Omega & = \cW_1 \, J \wedge J + \cW_2 \wedge J + \overline \cW_5 \wedge \Omega \ ,
\end{align}
with suitable torsion classes $\cW_1$, $\cW_2$, $\cW_3$, $\cW_4$, $\overline \cW_5$.
We can then see that in our case all torsion classes vanish, with the exception of
\beq
\overline \cW_5 = 0 + \alpha \, \overline \cW_5 ^\tbone
\ , \qquad
\overline \cW_5^\tbone = - 768 \, (2\pi)^3 e^{-\frac 32 \phi_0} \; \bar \partial^\tbzero Q \ .
\eeq
The corrected geometry is still K\"ahler and the manifold is an
almost Calabi-Yau threefold~\cite{Joyce:2001xt}.
The situation at hand can be compared to considering,
for instance, a quintic in $\mathbb P^4$ endowed with the 
metric induced by the ambient space Fubini-Study metric.
It is interesting to point out, however, that if the requirement
of simple connectedness is relaxed, it is possible to have
Ricci-flat K\"ahler  manifolds that are almost Calabi-Yau, but
not Calabi-Yau. An example is furnished by the Enriques
surfaces; recently, novel examples have been constructed in \cite{Andriot:2015sia}
in the context of solvmanifolds.

\subsubsection{Fluctuations associated to K\"ahler moduli}

In order to derive the relevant couplings in the four-dimensional
effective action of type IIB compactified on $Y_3$ we need to activate 
K\"ahler structure deformations of the internal metric entering 
the 10d background solution. We thus imagine to pick a fixed, reference point
in the complex structure and K\"ahler structure moduli space of $Y_3$
and to switch on small deformations in the K\"ahler structure moduli space directions.
 
It is well-known that, at zeroth-order in $\alpha$, the 
K\"ahler structure deformations of the 
internal Ricci-flat metric 
take the form
\beq \label{zeroth_deformation}
\delta g^\tbzero_{m \bar n} = - i \, \delta v^i \, \omega_{i m \bar n} \ ,  
\eeq
where $i = 1, \dots , h^{1,1}(Y_3)$,
$\delta v^i$ are real deformation parameters, and 
$\omega_{i m\bar n}$ denote a basis of harmonic $(1,1)$-forms
whose cohomology classes are Poincar\'e dual to 
an integral basis of the homology $H_4(Y_3, \mathbb Z)$.

At first order in $\alpha$ the structure of the internal metric deformations
can be written as
\beq 
\delta (g^\tbzero _{m \bar n} +  \alpha \, g^\tbone_{m \bar n} ) = 
- i \, \delta v^i \, (\omega_{i m\bar m} + \alpha \, \nabla^\tbzero_m \nabla^\tbzero_{\bar n}
\rho_i ) + \alpha \, \nabla^\tbzero_m \nabla^\tbzero_{\bar n} \delta \tilde F 
+ \cO(\alpha^2) \ .
\eeq
The $\rho_i$ functions parametrize deviations from the harmonic representative
within each cohomology class in $H^{1,1}(Y_3)$, while $\delta \tilde F$ denotes the 
variation of the function $\tilde F = *^\tbzero_6 (J^\tbzero \wedge F)$ 
appearing in  \eqref{gtbone_solution}
induced by the zeroth-order deformation \eqref{zeroth_deformation}.
Both kinds of modifications can be combined in a single expression of the form
\beq
\delta (g^\tbzero _{m \bar n} +  \alpha \, g^\tbone_{m \bar n} ) = 
- i \, \delta v^i \, (\omega_{i m\bar m} + \alpha \, \nabla^\tbzero_m \nabla^\tbzero_{\bar n}
\lambda_i )  + \cO(\alpha^2) \ ,
\eeq
for suitable functions $\lambda_i$. 
Similar results were found in \cite{Grimm:2014efa, Grimm:2015mua}. 
Crucially, these functions drop out from
the dimensional reduction discussed in the next section. As a consequence, we do not
need to discuss their specific form, and moreover it seems that they do not have any physical significance.
Let us point out that the deformation
\eqref{zeroth_deformation} of the zeroth-order metric 
induces also a modification of the quantity $Q$
entering the order $\alpha$ expressions for the Weyl rescaling function $\Phi^\tbone$
 and the dilaton
correction $\phi^\tbone$. Since we are working up to quadratic order in
fluctuations, we can consider a truncated Taylor series expansion 
for $Q$ of the form
\beq \label{Q_Taylor}
Q[g^\tbzero + \delta g^\tbzero] = Q + Q_{i} \delta v^i + \tfrac 12 Q_{ij} \delta v^i \delta v^j \ .
\eeq

In order to compute the 4d effective action we treat 
the fluctuation parameters $\delta v^i$ as arbitrary functions
of the external coordinates.
In a similar way, we promote the $\cO(\alpha^0)$
 dilaton profile $\phi_0$
from a constant to an arbitrary function of external spacetime.
In summary, 
the   Ansatz utilized in the dimensional reduction
takes the form
\begin{align} \label{full_Ansatz}
ds_{10}^2 \; \; & = \; \exp \left\{ -192  \, \alpha \, e^{-\frac 32  \phi(x) } \left(
 Q + Q_{i} \delta v^i + \tfrac 12 Q_{ij} \delta v^i \delta v^j 
 \right)  \right\} \times   \\
 \;\;\; & \;\;\; \times \bigg[
g_{\mu\nu} dx^\mu dx^\nu
+ 2 \Big(
g^\tbzero_{m \bar n} + \alpha \, g^\tbone_{m \bar n}
 - i \delta v^i (\omega_{i m \bar n } 
 + \alpha  \nabla^\tbzero_m \nabla^\tbzero_{\bar n} \lambda_i)
\Big ) dz^m d\bar z ^{\bar n}
  \bigg]
  + \cO(\alpha^2) + \cO(\delta v^3) \ , \nn \\
\wh \phi(x,y) & = \phi(x)   
+1152 \,    \tilde c_1 \, \alpha \, e^{-\frac 32  \phi (x)  } \,
\left( 
Q + Q_{i} \delta v^i + \tfrac 12 Q_{ij } \delta v^i \delta v^j
\right)
+ \cO(\alpha^2) + \cO(\delta v^3)  \ .
\end{align}
Note that we have replaced
the external Minkowski metric with an arbitrary metric $g_{\mu\nu}(x)$.
We will drop the explicit dependence of the spacetime  coordinates,
writing $\phi$ for $ \phi(x)$, for notational simplicity in the following.

\subsection{Reduction results}\label{redres}

This section is devoted to the discussion of the results of the dimensional
reduction of the various terms in \eqref{R4actiontot}
according to the Ansatz \eqref{full_Ansatz}.
In section \ref{results_uplift_section} we present the outcome of the computation and 
we address the problem of uplifting it from small fluctuations $\delta v^i$
to a full, non-linear dependence on the K\"ahler structure moduli.
Section \ref{weyl_rescaling_section} is then devoted to the Weyl rescaling that casts the 
4d Einstein-Hilbert term into canonical form.

\subsubsection{Results and uplift} \label{results_uplift_section}

To begin with, the reduction of the 
Einstein-Hilbert term to four dimensions yields
\beq
 \int_{\cM_{10}}  R \ast_{10} 1 \longrightarrow  
  \int_{\cM_4}  \left( \Omega \,  R   + 
P_{ij} \,  \nabla_\mu \d v^i  \nabla^\mu \d v^j 
 + P_i \, \nabla_\mu \delta v^i \nabla^\mu \phi
\right  )  *_4 1    \;\; ,
\eeq
with
\begin{align} \label{Omega_expr}
\Omega \; & = \int_{Y_3}    \Big[ 1  - i   \d v^i  \, \o _{i m}{}^{m}  +  \tfrac{1}{2}  \d v^i \d v^j  ( \o _{i m \bar n}  \o _{j }{}^{\bar n m} -  \o _{i m}{}^m  \o _{j n}{}^n )   - 384 \cdot 2  \, \alpha  \, e^{-\frac 32 \phi_0}  \, Q        
\Big] *^\tbzero_61 \ ,    \\
P_{ij} & = \int_{Y_3}   \Big[
\tfrac{1}{2}  \o_{i m \bar n}  \o_{j }{}^{\bar n m}  - \o_{i m}{}^m  \o_{j n}{}^n    - 384 \cdot 2 \, \alpha \, e^{-\frac 32 \phi_0} \, Q \, \left( 
\tfrac{1}{2}  \o_{i m \bar n}  \o_{j }{}^{\bar n m}  + \o_{i m}{}^m  \o_{j n}{}^n
\right)  
\Big ]  *^\tbzero_61 \ , \label{Pij_expr} \\
P_i \; & = \int_{Y_3} - 384 \cdot 6 \,   i \, \alpha  \, e^{-\frac 32 \phi}   \, \omega_{im}{}^m  \, Q
\, *^\tbzero_6 1 \ .  \label{Pi_expr}
\end{align}
The 
terms proportional to $\alpha$ originate from the 
10d Weyl rescaling factor in the backreacted metric Ansatz \eqref{full_Ansatz}.
  Let us point out that terms involving other quantities
such as    $g^\tbone_{m \bar n}$ and $\lambda_i$
drop out of the final result because their contributions
can be organized into total derivatives in the internal space.
It is also crucial to make use of the fact that the 
fluctuation of $c_3$ under a K\"ahler deformation
is an exact six-form, in accordance with $c_3$
being a characteristic class.\footnote{This observation allows us to
derive useful identities involving the variation of the zero-form $Q$.
Recalling $Q = (2\pi)^3 *^\tbzero_6 c_3$ and taking into account the variation
of the metric  implicit in the Hodge star, one can show that 
\beq
\int_{Y_3} \left[ Q_i   - i \, \omega_{im}{}^m \, Q\right] *^\tbzero_6 1 = 0 \ , \qquad
\int_{Y_3} \left[ Q_{ij}  +  
( \omega_{im \bar n} \, \omega_j{}^{\bar n m}
+ \omega_{im}{}^m \, \omega_{jn}{}^n ) Q   \right] *^\tbzero_6 1 = 0 \ ,
\eeq
where the quantities $Q_i$, $Q_{ij}$ were introduced in \eqref{Q_Taylor}.}

Next we consider the reduction of the dilaton kinetic terms.
Note that $\nabla_\mu \phi$ can be effectively considered 
to be a fluctuation of the same order as $\delta v^i$, and as a result
terms of the schematic form $\delta v \nabla \phi \nabla \phi$
or $\delta v \nabla \delta v \nabla \phi$
have to be neglected.
To linear order in fluctuations we have
\beq
\nabla_\mu \phi = \nabla_\mu   \phi \, \left[
1 - 1728   \, \alpha \, \tilde c_1  \, e^{-\frac 32  \phi  }
Q 
\right] 
+ 1154 \,  \alpha \, \tilde c_1  \, e^{-\frac 32 \phi} \, Q_i \, \nabla_\mu \delta v^i  \ .
\eeq
Combining this observation with the effect of the Weyl rescaling factor we obtain
\begin{align}
 \int _{\cM_{10}} - \tfrac{1}{2} \nabla_M \wh \phi \nabla^M \wh \phi \ast_{10} 1 
 \longrightarrow \; \; &
\int_{\cM_4}  \left[ 
U \, 
\nabla_\mu  \phi \nabla^\mu   \phi
+ U_i \, \nabla_\mu \phi \nabla ^\mu \delta v^i
 \right]
 \, *_41 \ ,  
 \end{align}
 with
\begin{align}
U    = -  \tfrac 12  (2\pi \alpha')^3 \, \cV^\tbzero  + 384 \, \alpha  (1+ \tfrac 92  \,  \tilde c_1 ) 
e^{-\frac 32 \phi} \, (2\pi)^3 \chi \ , \qquad
U_i  = - 384 \alpha \cdot 3 i \, \tilde c_1 \, \omega_{im}{}^m\, (2\pi)^3 \chi  \ , 
\end{align}
where $\cV^\tbzero$ denotes the volume of $Y_3$
in units of $\sqrt{2\pi \alpha'}$ computed with the
zeroth-order metric $g^\tbzero_{m \bar n}$ 
and we used \eqref{Q_integral} and the footnote   for the integrals of $Q$,
$Q_i$. We have also recalled that $\omega_{im}{}^m$ is
constant on the threefold.

We can now record the results of the reduction of the higher-derivative
terms.
First of all, 
\beq
\int_{\cM_{10}} e^{-\frac 32 \phi} \left( t_8 t_8 + \frac 18 \epsilon_{10} \epsilon_{10}\right) R^4 \, *_{10} 1
\longrightarrow  
\int_{\cM_4} 
\left[
 - 384 \cdot 2  (2\pi)^3 e^{-\frac 32 \phi } \chi \, R + \tilde P_{ij} \, \nabla_\mu \delta v^i \nabla^\mu \delta v^j
 \right] *_41 \ ,
\eeq
where   
\begin{align} \label{tildePij_expr}
\tilde P_{ij} & =  384 (2\pi)^3  e^{-\frac 32 \phi } \int_{Y_3} \omega_{im \bar n} {\omega_j}^{\bar n m} \, c_3 \ .
\end{align}
Let us remind the reader that the internal six-form $c_3$ is the third Chern form defined in appendix \ref{Conv_appendix}
and was related to $Q$ in \eqref{Q_integral}.
Secondly,
the reduction of the higher-derivative dilaton
couplings in \eqref{R4actiontot} yields
\ba
&\int_{\cM_{10}}  
e^{- \frac 32 \phi}
\bigg(  
\tilde c_1 \nabla^M  \nabla_M \wh \phi \, f_1
+ \tilde c_2  \, \nabla_M \wh \phi \, \nabla_N \wh \phi \, f_2^{MN}
+ \tilde c_3  \, \nabla^M \wh \phi \, \nabla_M \wh \phi \, f_3
\bigg)   \\ &\quad  \quad 
\longrightarrow  \quad 384 (2\pi  )^3
\int_{\cM_4} 
e^{-\frac 32 \phi}
\bigg[
- \tfrac 32 (3 \tilde c_1 + 2 \tilde c_2 +4 \tilde c_3) \, \chi \, 
\nabla_\mu   \phi \nabla^\mu   \phi
+  3 i \, \tilde c_1 \, \omega_{im}{}^m  \,\chi \,  \nabla_\mu \phi \nabla^\mu \delta v^i 
\bigg]    *_41 \ . \nn
\ea

As mentioned above, in the process of dimensional reduction 
we have implicitly chosen a reference point in
the complex structure and K\"ahler moduli spaces of the threefold
$Y_3$ and we have only activated K\"ahler fluctuations $\delta v^i$,
retaining  terms up to quadratic order.
Our next task is to infer the form taken by the couplings written above
when the fluctuations $\delta v^i$ are promoted to a full
dependence on the moduli space of $Y_3$.
At two-derivative level this is a standard procedure 
that we briefly
review in order to set up our notation.

Recall that the $(1,1)$ forms $\omega_{im\bar n}$,
$i = 1, \dots, h^{1,1}$ are the 
harmonic representatives of the cohomology classes
dual to an integral basis $D_i$ of the homology  
$H_4(Y_3 , \mathbb Z)$.
The threefold intersection numbers are denoted
\beq
\cK_{ijk} = D_i \cdot D_j \cdot D_k  =(2\pi \alpha')^{-3}\, \int_{Y_3} 
\omega_i \wedge \omega_j \wedge \omega_k \ .
\eeq
We have inserted the appropriate power of $2\pi \alpha'$
in order to make the intersection numbers $\cK_{ijk}$ dimensionless.
Indeed,   the components $\omega_{i m \bar n}$ are dimensionless,
so that the $(1,1)$-forms $\omega_i = \omega_{i m \bar n} \, dz^m \wedge d\bar z ^{\bar n}$
have mass dimension $-2$.

The K\"ahler class 
of the zeroth-order metric at the reference point in 
moduli space 
can be expanded onto the basis of forms $\omega_i$ as
\beq \label{J_expansion}
J^\tbzero = v^\tbzero{}^i \omega_i \ ,
\eeq
where   $v^\tbzero{}^i$  are taken to be
dimensionless.
We also define the quantities
\beq \label{curlyKquantities}
\cK_{ij}^\tbzero = \cK_{ijk} v^\tbzero{}^k  \ , \qquad
\cK_{i}^\tbzero = \tfrac 12 \cK_{ijk} v^\tbzero{}^j v^\tbzero{}^k \ , \qquad
\cV^\tbzero = \tfrac16 \cK_{jik} v^\tbzero{}^i v^\tbzero{}^j v^\tbzero{}^k \ . 
\eeq
The quantity $\cV^\tbzero$ is
the volume of the threefold in units of $\sqrt{2\pi \alpha'}$,
so that we can write
\beq
 \int_{Y_3} *^\tbzero_6 1 =  (2\pi \alpha')^3\, \cV^\tbzero \ .
\eeq
Let us record the useful identities
\begin{align} \label{topologicalPij}
\omega_{im}{}^m & = i \frac{\cK^\tbzero_i}{\cV^\tbzero} \ , &
\omega_{im \bar n} \omega_j{}^{\bar nm} \, *^\tbzero_61 & = 
\omega_i \wedge \omega_j \wedge J - \frac{\cK^\tbzero_i \cK^\tbzero_j}{\cV^\tbzero{}^2} \,
 *^\tbzero_61 \ ,
\end{align}
and stress that harmonicity of $\omega_i$ guarantees that $\omega_{im}{}^m$
is covariantly constant on $Y_3$. 
Using these identities we can immediately compute
the $\cO(\alpha^0)$ part of $P_{ij}$ defined in \eqref{Pij_expr},
\beq
P_{ij} =(2\pi \alpha')^3 \left[
 \tfrac 12 \cK^\tbzero_{ij} + \tfrac 12 \tfrac{\cK^\tbzero_i \cK^\tbzero_j}{\cV^\tbzero}
 \right] + \cO(\alpha) \ .
\eeq
This expression is initially understood to be evaluated at the reference point in
K\"ahler moduli space, but given its topological nature it uplifts naturally
to the full moduli space dependence.
This is achieved simply by replacing $v^\tbzero{}^i$ with an arbitrary $v^i$.
The quantities defined in \eqref{J_expansion}, \eqref{curlyKquantities}
in terms of $v^\tbzero{}^i$ are promoted to $v$-dependent
quantities denoted $\cV$, $\cK_i$, $\cK_{ij}$ without the ${}^\tbzero$ superscript.
In a similar way,
the quantity $\Omega'$ naturally uplifts to the full $v$-dependent threefold volume 
form, which integrates to the full volume.

The uplift of higher-derivative couplings is considerably less under control in general,
due to the fact that corrections are expected to lift some of the na\"ive flat directions
of moduli space, making the identification of the correct massless modes a hard
problem. Nonetheless, we can recast 
the reduction results of the higher-derivative terms under consideration
in a simple form. 
Exploiting the fact that $\omega_{im}{}^m$ is covariantly constant we can
immediately perform the uplift of  integrals of the form
\beq
\int_{Y_3} \omega_{im}{}^m \, c_3 \longrightarrow i\, \chi \, \frac{\cK_i}{\cV} \ , \qquad
\int_{Y_3} \omega_{im}{}^m \, \omega_{jn}{}^n \, c_3 \longrightarrow - \chi \frac{\cK_i \cK_j}{\cV^2} \ .
\eeq
A similar manipulation for the integral
\beq
\int_{Y_3} \omega_{im\bar n} \, \omega_{j}{}^{\bar n m} \, c_3
\eeq
is not straightforward, since the non-harmonic part $F$ of $c_3$
introduced in \eqref{deldelbarconsequence} poses an obstruction to the factorization of this expression. 
In the present context, however, the overall coefficient of this term in the dimensional
reduction is zero and we do not have to address this complication.

Let us introduce the field-dependent, dimensionless 
quantity 
\beq   \label{tildechi_expression}
\tilde \chi = 384 (2\pi)^3 \, \frac{\alpha}{(2\pi \alpha')^3 \, \cV} \, e^{-\tfrac 32 \phi} \, \chi
=     \tfrac {1}{16} \zeta(3) \frac{e^{-\tfrac 32 \phi}}{\cV}   \, \chi \ .
\eeq
In what follows it plays the role of the effective expansion parameter
for the problem at hand.
Combining all contributions after uplift
the 4d effective action takes the form
\begin{align} \label{beforeWeyl}
S_{\rm 4d} =\tfrac{1}{(2\pi )^4\alpha'} \int_{\cM_4} \bigg\{ &  \cV  \left( 1+ b_1 \, \tilde \chi   \right) R
+ \bigg[   
\left( \frac 12 + b_{2} \, \tilde \chi \right) \cK_{ij}
+ \left( \frac 12 + b_{3} \, \tilde \chi\right) \frac{\cK_i \cK_j}{\cV}
 \bigg]   
  \nabla_\mu v^i \nabla^\mu v^j   \nn \\
& + \cV 
 \left(
 - \frac 12  + b_4 \, \tilde \chi  
 \right) 
 \nabla_\mu \phi \nabla^\mu \phi
 +  b_5 \, \tilde \chi \, \cK_i \, \nabla_\mu \phi \nabla^\mu v^i
 \bigg\} *_4 1  \ .
\end{align}
In this expression we have promoted the fluctuations $\delta v^i$ 
to full 4d fields $v^i$. For notational convenience we have introduced the numerical coefficients
\begin{align} \label{b_coefficients}
b_1 & = -4 \ , &
b_{2} & =0 \ , &
b_{3} & = 2  \ , \\
b_4 & = 1    
- 3 \tilde c_2 - 6 \tilde c_3\ , &
b_5 & = 6 \ .
\end{align}

\subsubsection{Weyl rescaling } \label{weyl_rescaling_section}

Our next task is performing a Weyl rescaling of the external metric in \eqref{beforeWeyl}
in order to bring the 4d Einstein-Hilbert term into  canonical form.
Let us record the identities
\beq
g_{\mu\nu} = e^{2\Lambda} \tilde g_{\mu\nu} \ , \qquad
R = e^{- 2 \Lambda} \left[ \tilde R - 6 \tilde \nabla_\mu \Lambda \tilde \nabla^\mu \Lambda 
- 6 \tilde \nabla_\mu \tilde \nabla^\mu \Lambda \right] \ .
\eeq
In our case, we set
\beq
\Lambda = - \tfrac 12 \log \left[ \cV \left(  1+ b_1 \, \tilde \chi \right)\right] \ .
\eeq
Dropping the tilde on the new metric,
the Weyl-rescaled 4d action 
can be written in the form
\beq \label{afterWeyl}
S_{\rm 4d} =\tfrac{1}{(2\pi )^4\alpha'} \int_{\cM_4}  \bigg[  R   + \cG_{v^i v^j} \, \nabla_\mu v^i \nabla^\mu v^j
+ \cG_{\phi \phi} \, \nabla_\mu \phi \nabla^\mu \phi
+ 2 \, \cG_{\phi v^i} \, \nabla_\mu \phi \nabla^\mu v^i
\bigg] *_41 \ ,
\eeq
where  
\begin{align} \label{vMetric}
\cG_{v^i v^j} &  =- \frac{1}{\cV^2} \left[ 1
 + \tilde \chi  \, 
 \left(-b_{3} - \tfrac 52 b_1 \right)
 \right]\,
 \cK_i \cK_j
+ \frac{1}{2\cV} \left[ 1  
+ \tilde \chi \, (2b_{2} - b_1)
  \right] \, \cK_{ij} \ ,   \\
\cG_{\phi \phi} & = - \frac{1}{2} \left[
1 
+ \tilde \chi    \,
(-b_1 -2b_4)
 \right] \ ,   \\
 \cG_{\phi v^i} & =\left(
  \tfrac 94 b_1  + \tfrac 12 b_5
  \right) \tilde \chi  \frac{\cK_i}{\cV}
 \ .
\end{align}
This form of the reduction result
is a convenient starting point for 
the discussion 
of the 4d $\cN = 2$ and 
 $\cN = 1$ theories in the following sections.

\section{Correction to the $\cN = 2$ prepotential} \label{N2section}

In this section we analyze the result of the dimensional reduction
from the point of view of the 4d, $\cN = 2$   effective theory
obtained without including branes or performing any orientifold projection.
We show that our findings are compatible with the 
known results about the
correction to the geometry of the 
$\cN = 2$ hypermultiplet scalar manifold
induced by a perturbative $\alpha'$ correction to 
 the K\"ahler moduli space prepotential
\cite{Candelas:1990rm, Candelas:1990qd,Antoniadis:1997eg}.

\subsection{Translation into $\cN = 2$ field variables}

Let us recall some well-known facts about the 4d, $\cN = 2$ effective
theory arising from compactification of type IIB superstring on a Calabi-Yau
threefold \cite{Ferrara:1988ff, Ferrara:1989ik, Bodner:1989cg,  Bohm:1999uk}. The complex structure moduli of the threefold
fit into $n_V= h^{1,2}$   vector multiplets. Their scalar manifold is special K\"ahler and its
geometry is tree-level exact both in $g_s$ and $\alpha'$.
The  K\"ahler structure moduli, the dilaton, the axion, as well as the scalars
coming from the expansion of $B_2$, $C_2$, $C_4$
onto internal even harmonic forms,
all fit into $n_H = h^{1,1} +1$ hypermultiplets.
Their scalar manifold $\cM_{\rm Q}$ is quaternionic of real dimension
$4(h^{1,1} + 1)$.

The structure of the metric on $\cM_{\rm Q}$
is most conveniently analyzed using the string frame K\"ahler moduli
$v^i_{\rm s}$ 
and the 4d   dilaton $\phi_4$. 
In terms of these variables the quaternionic metric on $\cM_{\rm Q}$ can be 
written schematically as \cite{Ferrara:1988ff, Ferrara:1989ik}
\beq \label{quaternionic_metric}
ds^2 (\cM_{\rm Q}) = (d\phi_4)^2 + G_{i \bar \jmath}(w_{\rm s}, \bar w_{\rm s}) \, dw^i \, d\bar w^{\bar \jmath}
+ \cG_{xy}(w_{\rm s}, \bar w_{\rm s}, \phi_4, q) dq^x \, dq^y \ .
\eeq
In this expression $w^i_{\rm s} = u^i_{\rm s} + i \, v^i_{\rm s}$ are the 
complexified K\"ahler moduli arising from expansion of $B_2 + i J_{\rm s}$
onto the basis $\omega_i$ of harmonic $(1,1)$-forms.
The metric $G_{i \bar \jmath}$ is the special K\"ahler
metric on the K\"ahler moduli space $\cM_{\rm K}$ of the threefold.
The fields $q^x$ denote collectively all real scalars in $\cM_{\rm Q}$
different from $\phi_4$, $u^i_{\rm s}$, $v^i_{\rm s}$.
The metric components $\cG_{xy}$
are entirely determined by the special K\"ahler metric
on $\cM_{\rm K}$. For this reason, the quaternionic
metric on $\cM_{\rm Q}$ is sometimes referred to as special quaternionic \cite{Ferrara:1988ff, Ferrara:1989ik}.
Let us stress that
the structure \eqref{quaternionic_metric} of the metric on $\cM_{\rm Q}$
is expected to hold not only at tree level,
but also including $\alpha'$ corrections,
thanks to mirror symmetry considerations,
reviewed for instance in \cite{Hosono:1994av, Hori:2003ic}.

In our analysis all the fields $q^x$ as well as the axions $u^i_{\rm s}$ are effectively frozen to zero.
Our next task is therefore to connect 
the string frame K\"ahler moduli $v^i_{\rm s}$
and the 4d dilaton $\phi_4$
to the 
Einstein frame K\"ahler moduli $v^i$ and the 10d dilaton $\phi$.
At two-derivative level we have the relations
\beq
e^{-2 \phi_4} = e^{- \frac 12 \phi} \, \cV \  , \qquad
v_{\rm s}^ i = e^{\frac 12 \phi} \, v^i  \ .
\eeq
Let us remind the reader that in our conventions $\cV$
is the volume of the threefold in units of $\sqrt{2\pi \alpha'}$.
In order to take into account the effect of higher-derivative corrections
we deform these relations into
\beq \label{deformed_string_vars}
e^{-2 \phi_4} = e^{- \frac 12 \phi} \, \cV \, \left( 1 + \Upsilon_1 \, \tilde \chi \right) \  , \qquad
v_{\rm s}^ i = e^{\frac 12 \phi} \, v^i \, \, \left( 1 + \Upsilon_2 \, \tilde \chi    \right)   \ ,
\eeq
where 
  $\Upsilon_1$, $\Upsilon_2$ are constants that will be fixed momentarily.

We can now invert   \eqref{deformed_string_vars}
to leading order in $\alpha'^3$ to obtain $v^i$, $\phi$
in terms of $v^i_{\rm s}$, $\phi_4$, and plug the resulting
expressions into the 4d effective action \eqref{afterWeyl}.
The result is conveniently expressed in terms of the quantities
\beq
\cV_{\rm s} = \tfrac 16 \cK_{ijk} \, v^i_{\rm s} \, v^j_{\rm s} \, v^k_{\rm s}
\ , \qquad
\cK^{\rm s}_i = \tfrac 12 \cK_{ijk} \,  v^j_{\rm s} \, v^k_{\rm s}  \ , \qquad
\cK^{\rm s}_{ij} =   \cK_{ijk} \,    v^k_{\rm s}   \ .
\eeq
Furthermore, it is convenient to 
introduce a new field-dependent dimensionless quantity $\tilde \chi_{\rm s}$,
which is the string-frame analog of 
$\tilde \chi$ and is given by
\beq
\tilde \chi_{\rm s} =    \tfrac {1}{16} \zeta(3) \frac{1 }{\cV_{\rm s}}\,  \chi
 \ .
\eeq
With this notation, the 4d effective action can be written as  
\begin{align} \label{string_variable_action}
S_{\rm 4d}  =\tfrac{1}{(2\pi)^4 \alpha'} \int _{\cM_4 }\bigg\{ & R  + \left[
 \frac{1}{2 \cV_{\rm s}} \left(1 +   a_1 \, \tilde \chi_{\rm s} \right)
\cK^{\rm s}_{ij}
- \frac{1}{2 \cV_{\rm s}} \left(
1 +   a_2 \, \tilde \chi_{\rm s}
 \right) \frac{\cK_i^{\rm s}  \cK_j^{\rm s}}{\cV_{\rm s}}
\right] \nabla_\mu v^i_{\rm s} \nabla^\mu v^j_{\rm s} \nn \\
& - 2 \left( 1 +   a_3 \, \tilde \chi_{\rm s} \right)
\nabla_\mu \phi_4 \nabla^\mu \phi_4
+   a_4 \, \tilde \chi_{\rm s} \, \frac{\cK_i^{\rm s}}{\cV_{\rm s}} \, \nabla_\mu \phi_4 \nabla^\mu v^i_{\rm s}
\bigg\} *_41 \ ,
\end{align}
where the numerical coefficients $  a$ are given
in terms of the $b$ coefficients introduced in  \eqref{beforeWeyl} by
\begin{align}
  a_1 & = 2b_2 - b_1 \ , & 
  a_2 & = - \tfrac{37}{16} b_1  + \tfrac 54 b_2 - \tfrac 18 b_3 - \tfrac 12 b_4 -\tfrac 14 b_5 
+ 2 \Upsilon_2\ , \nn \\
  a_3 & = \tfrac{11}{16} b_1 - \tfrac 34 b_2 - \tfrac 98 b_3 - \tfrac 12 b_4+ \tfrac 34 b_5 \ , &
  a_4 & = -\tfrac{27}{8} b_1- \tfrac 12 b_2 - \tfrac 34 b_3 + b_4 - \tfrac 12 b_5
+ 2 \Upsilon_1 \ .
\end{align}
If we plug in the values of the $b$ coefficients given in \eqref{b_coefficients}, we find
\begin{align}
  a_1 & =4 \ , & 
  a_2 & =7 + \tfrac 32 \tilde c_2 + 3 \tilde c_3
+ 2 \Upsilon_2\ , \nn \\
  a_3 & = - 1 + \tfrac 32 \tilde c_2 + 3 \tilde c_3 \ , &
  a_4 & = 13 - 3 \tilde c_2 - 6 \tilde c_3
+ 2 \Upsilon_1 \ .
\end{align}
We are now in a position to discuss the correction to the $\cN = 2$ prepotential.

\subsection{Implications for $\cN =2$ prepotential} \label{implications_for_F}

As noted above, the form \eqref{quaternionic_metric}
of the quaternionic metric on $\cM_{\rm Q}$
should be preserved by $\alpha'$ corrections. In particular,
cross terms between $\phi_4$ and $v^i_{\rm s}$ are not allowed,
and also $v^i_{\rm s}$-dependent corrections to the $(d\phi_4)^2$
terms are forbidden. These considerations lead us to impose
$  a_3 = 0$, $  a_4 = 0$. As a result we can fix 
the value of $\Upsilon_1$ 
and derive a linear constraint on the  
dilaton coupling coefficients $\tilde c_2$, $\tilde c_3$,
\beq \label{cross_terms_constraints}
\Upsilon_1 = -4 \ , \qquad
3 \tilde c_2  + 6 \tilde c_3 - 2= 0 \ .
\eeq

We now aim at demonstrating that
the effective action \eqref{string_variable_action}
can be written as
\beq
S = \tfrac{1}{(2\pi)^4 \alpha'} \int_{\cM_4} \bigg\{
R - 2 \nabla_\mu \phi_4 \nabla^\mu \phi_4
- 2  G_{i \bar \jmath}\, \nabla_\mu v^i_{\rm s}  \nabla^\mu v^j_{\rm s}
\bigg\} *_4 1 \ , \qquad 
G_{i \bar \jmath} = \partial_{w^i_{\rm s}} \partial_{\bar w^{\bar \jmath}_{\rm s}} K^{\cN = 2}(w_{\rm s} , \bar w_{\rm s}) \ ,
\eeq
where we remind the reader that $w^i_{\rm s} = u^i_{\rm s} + i \, v^i_{\rm s}$ are the complexified
K\"ahler moduli, and where the K\"ahler potential $K^{\cN = 2}( w_{\rm s} , \bar w_{\rm s})$
can be derived from a holomorphic prepotential.

In order to set our notation, it is useful to
review the relation between the K\"aher potential
and the prepotential. The geometry of the
 $h^{1,1}$ complex-dimensional K\"ahler moduli space
 is conveniently described 
by $h^{1,1} +1$ complex projective coordinates $X^I = (X^0, X^i)$
related to
the complex coordinates $w^i_{\rm s}$ by
\beq
w^i_{\rm s} = \frac{X^i}{X^0} \ .
\eeq
The geometry of $\cM_{\rm K}$ is   encoded in a holomorphic prepotential
$F(X)$, which is a homogeneous function of $X^I$ of degree 2.
The K\"ahler potential $K^{\cN = 2}(w_{\rm s} , \bar w_{\rm s})$ for the metric on $\cM_{\rm K}$ 
is then extracted from $F$ via
\beq \label{from_F_to_K}
e^{- K^{\cN = 2}( w_{\rm s} , \bar w_{\rm s})}
= -i( X^I \overline F_I - F_I \overline X^I)  \ , \qquad
F_I = \partial_{X^I} F(X) \ .
\eeq
Mirror symmetry considerations \cite{Candelas:1990rm,Hosono:1994av,Hori:2003ic} ensure that the prepotential $F(X)$
takes the form
\beq \label{full_prepotential}
F(X) = \frac{1}{X^0} \frac 16 \cK_{ijk} X^i X^j X^k  + i \, \lambda \, (X^0)^2 + \dots\ ,
\eeq
where the first term is the classical prepotential,
the second is a perturbative $\alpha'$ effect, and 
the ellipsis stands for terms that are exponentially suppressed
in the large volume limit and originate from worldsheet instantons.
The $\cN = 2$ K\"ahler potential derived from 
\eqref{full_prepotential} using \eqref{from_F_to_K} is, up to instanton corrections 
and an immaterial constant,
\beq \label{N2Kaehler}
K^{\cN = 2} = - \log \left[ \cV_{\rm s} + \tfrac 12 \lambda  \right] \ .
\eeq

We can finally relate this to our dimensional reduction
by comparing the metric computed from \eqref{N2Kaehler}
with the metric in \eqref{string_variable_action}.
We find that, in order for the match to be possible,
we have to impose
\beq
\Upsilon_2 = 0 \ .
\eeq
The constant $\lambda$ is then fixed to be
\beq
\lambda 
= - \tfrac {1}{2} \,  \zeta(3) \, \chi  \ ,
\eeq
so that the 
 $\cN = 2$ K\"ahler potential
 reads
\beq \label{N2_potential}
K^{\cN =2} = 
- \log \left[ \frac{\cV_{\rm s}}{(2\pi \alpha')^3} -   \tfrac 14   \zeta(3)  \,  \chi   \right] \ .
\eeq
In this last expression we have reinstated all factors of $\alpha'$
and $\cV_{\rm s}$ denotes the dimensionful volume of the threefold.
Let us close this section
with a comparison between our findings and the analogous
quantities in BBHL \cite{Becker:2002nn}.
Using equations (3.11), (3.12), (3.13) 
and the comment before (3.15) in that paper,
we infer that their 
$\cN = 2$ K\"ahler potential in our notation
coincides exactly with \eqref{N2_potential}. 


\section{$\cN = 1$ K\"ahler coordinates and K\"ahler potential} \label{Kaehler_section}

In this section we analyze the results of the dimensional reduction
and identify the K\"ahler coordinates and K\"ahler potential 
in the 4d, $\cN = 1$ effective action. We briefly comment  
on our findings.

\subsection{Correction to the K\"ahler potential and coordinates}

Upon dimensional reduction on a Calabi-Yau threefold, the two-derivative action
of type IIB supergravity yields a 4d effective action with $\cN = 2$ supersymmetry.
If  orientifold planes are included in the setup, the 4d spectrum
is suitably projected and one obtains an effective action with 4d, $\cN = 1$
supersymmetry \cite{Grimm:2004uq, Grimm:2004ua, Grimm:2005fa}. 
For definiteness, we consider here the projection relevant to the case
with $O3/O7$-planes.
The dilaton $\phi$ and the K\"ahler moduli $v^i$ fit into 4d, $\cN = 1$
chiral multiplets whose scalar components are of the form
\begin{align} \label{zerothCoordinates}
\tau_0 & = C_0 + i e^{-\phi}  \ , \\
T_i &= \rho_i + i\, \cK_i   + \zeta_i \ .
\end{align}
Several remarks are in order. We used the symbol $\tau_0$ to denote the 
4d axio-dilaton. Its real part $C_0$ is the straightforward dimensional
reduction of the fluctuations of the 10d axion around its zero background.
Its imaginary part is built with $\phi$ which is not directly identified
with the 10d dilaton because of the backreaction term $\phi^\tbone$
discussed in section \ref{background_without_fluctuations}. This discrepancy
has  consequences for the transformation property of 
4d fields under the 10d $SL(2,\mathbb Z)$ symmetry,
as discussed in \cite{Becker:2002nn}.
In the expression for $T_i$, 
 the scalars $\rho_i$ come from the expansion of the RR four-form,
and $\zeta_i$ denotes a complex quantity built with the scalars
coming from the reduction of the NSNS and RR two-forms.
Since we are freezing all such scalars to zero, 
the quantity $\zeta_i$ vanishes.
Let us point out that 
the orientifold projection to $\cN =1$ supersymmetry is
implicit in the range of the index $i$.
Indeed, by an abuse of notation,
 the index $i$ 
in \eqref{zerothCoordinates} 
runs only over the elements of $H^{1,1}_+(Y_3)$,
i.e.~the $(1,1)$ forms that are even under the isometric involution
of $Y_3$
considered in the implementation of the orientifold projection.
Let us stress   that equations \eqref{curlyKquantities} still hold 
after restricting the range of $i$, by virtue of the restrictions
imposed on intersection numbers by the orientifold projection,
as explained in detail in \cite{Grimm:2004uq, Grimm:2004ua, Grimm:2005fa}.

The kinetic terms of $\tau_0$ and $T_i$
are governed by a K\"ahler potential $K$.
If we write $T_0 \equiv \tau_0$ and introduce the collective
index $I = (0,i)$, the effective action contains the terms
\beq
S_{\rm 4d} \supset \tfrac{1}{(2\pi)^4 \alpha'}   \int _{\cM_4}
\left[
R + G_{T_I \overline T_J} \, \nabla_\mu T_I \nabla^\mu \overline T_J
 \right] *_4 1 \ ,
\eeq 
where
\beq
G_{T_I \overline T_J } = -2\,  \partial_{T_I} \partial_{\overline T_J} K \ , \qquad
K  = \phi - 2 \log \cV\ .
\eeq
The K\"ahler potential is understood 
as a function of $T_I$, $\overline T_J$
determined implicitly via \eqref{zerothCoordinates}.

Upon inclusion of higher-derivative terms in the 10d action,
the coefficients of the terms
in the two-derivative 4d effective action  
are modified, but the form of the action
is still expected to obey the constraints deriving from
4d,    $\cN = 1$ supersymmetry.
In this section we show that our results are consistent with 
this expectation.  

Let us introduce the notation $v^0 \equiv \phi$,
in such a way that $v^I = (\phi , v^i)$. 
The kinetic terms specified by \eqref{vMetric}
can be written compactly as
\beq
\cG_{v^I v^J} \, \nabla_\mu v^I \nabla^\mu v^J = 
\cG_{\phi \phi} \, \nabla_\mu \phi \nabla^\mu \phi
+ \cG_{v^i v^j} \, \nabla_\mu v^i \nabla^\mu v^j 
+ 2 \cG_{v^i \phi } \, \nabla_\mu v^i \nabla^\mu \phi \ .
\eeq
Recall 
 that $C_0$, $\rho_i$ and $\zeta_i$ are frozen to zero in our discussion.
The problem at hand is the determination of the $\cO(\tilde \chi)$
corrections to $T_I$ and $K$ in such a way that the relation 
 \beq
\cG_{v^I v^J} \, \nabla_\mu v^I \nabla^\mu v^J 
=-2 \, (\partial_{T_I} \partial_{\overline T_J} K )\,
\nabla_\mu T_I \nabla^\mu \overline T_J 
\eeq
holds including terms up to order  $\tilde \chi$.

The first outcome of our analysis is the observation that,
regarding the $b$ coefficients in \eqref{beforeWeyl} as input data,
 a solution exists only if the following linear constraint is 
 satisfied,
 \beq
11 b_1 - 12 b_2  - 18 b_3 - 8 b_4 + 12 b_5 = 0  \ .
\eeq
Plugging in the values of the $b$ coefficients given in \eqref{b_coefficients},
we obtain
\beq \label{linear_relation}
3 \tilde c_2  + 6 \tilde c_3 - 2= 0  \ ,
\eeq
which is the same constraint on $\tilde c_2$, $\tilde c_3$
found in \eqref{cross_terms_constraints} in the analysis
of the $\cN = 2$ case.
If this requirement is met, the K\"ahler coordinates 
and K\"ahler potential are   given in terms of the $b$ coefficients by
\begin{align}
\tau_0 & = i e^{-\phi} \left[1 +  \tilde \chi   \, 
\left(- \tfrac 34 b_1  + \tfrac 32 b_{3}  - b_5  \right)
\right] \ , \\
T_i & = i\, \cK_i \left[ 
1 +  \tilde \chi    \, 
\left( \tfrac 14  b_1 + 2 b_{2} + \tfrac 12 b_{3} \right)
\right] \ ,   \\
K & =  \phi -2 \log \left[ \cV \left( 
1 + \tilde \chi    \, 
\left(
b_1 + b_{2} + \tfrac 32 b_{3}  - \tfrac 12 b_5
  \right)
  \right)
\right] \ . \label{Kaeler_potential_bcoeff}
\end{align}
The actual values taken by these quantities upon substituting
the numerical values of the $b$ coefficients are summarized at the end of this section.

An important feature of the zeroth-order K\"ahler 
potential is the property of extended no-scale structure
 \cite{Grimm:2004uq, Grimm:2004ua, Grimm:2005fa}.
This property is tested by computing the following quantity,
\beq \label{extended_noscale}
\frac{\partial K}{  \partial T_I} \, (K^{-1})_{I \bar J } \, \frac{\partial K}{\partial \overline T_{\bar J}}
= 4 \left[ 
1 + 0 \cdot   \tilde \chi + \cO(\tilde \chi^2)    
\right] \ ,
\eeq
where $(K^{-1})_{I \bar J }$ denotes the inverse of the matrix
$K^{I \bar J} = \partial_{T_I} \partial_{\overline T_J} K$.
The zeroth order value $4$ signals extended no-scale structure 
\cite{Ciupke:2015ora},
and interestingly the leading correction has a vanishing coefficient
for any value of the $b$ coefficients.
This does not mean that the correction under examination has no physical
effect.
For instance, in a scenario like KKLT the axio-dilaton and 
the complex structure moduli are stabilized first at a supersymmetric
value by means of the Gukov-Vafa-Witten superpotential \cite{Kachru:2003aw}.
The relevant quantity in the computation of the scalar
potential for the K\"ahler moduli is no longer the quantity in \eqref{extended_noscale},
but rather the same object with summation restricted
to $i$, $j$ indices only, implicitly evaluated at the fixed value 
of the axio-dilaton. One finds
\beq
\frac{\partial K}{  \partial T_i} \, (K^{-1})_{i \bar \jmath } \, \frac{\partial K}{\partial \overline T_{\bar \jmath}}
= 3 \left[ 1  +  (\tfrac 12 b_1 - b_2) \, \tilde \chi + \cO(\tilde \chi^2)\right] \ .
\eeq
As we can see, the zeroth-order value $3$ associated to 
no-scale structure receives a non-zero correction
 at order $\tilde \chi$ ($\tfrac 12 b_1 - b_2 =- 2$).

In order to elucidate further the physical effects of the
correction to the K\"ahler potential
an alternative formulation can be used, in which the chiral multiplets
$T_i$ are dualized into linear multiplets $L_i$ \cite{Grimm:2004uq, Grimm:2004ua, Grimm:2005fa}.
The dynamics of the system is encoded in a kinetic potential $\tilde K$, which is the 
Legendre transform of the original K\"ahler potential $K$.
Let us adopt the following conventions for the Legendre transform,
\beq
L^i = - \frac{\partial K}{\partial {\rm Im} T_i} \ , \quad
\tilde K = K + L^i \,  {\rm Im} T_i \ , \quad
{\rm Im}T_i = \frac{\partial \tilde K}{\partial L^i} \ .
\eeq
A straightforward computation gives, up to terms
of order $\tilde \chi^2$ and higher,
\begin{align}
L_i & = \frac{v^i}{\cK} \left[
1 +  \tilde \chi    \,
\left(  -\tfrac 54 b_1 - \tfrac 12 b_{3} \right)
\right] \ , \\
L^i \, {\rm Im}T_i & = 3 \left[
1 +  \tilde \chi   \, 
\left(
-b_1  + 2 b_{2}
 \right)
 \right] \ . 
\end{align}
In this formulation the hallmark of zeroth-order   no-scale structure
is the value $3$, which again receives corrections at order $\tilde \chi$.
As a further check that the 4d dynamics is really affected by the
corrections under examination we can examine the expression
for the kinetic potential $\tilde K$ in terms of the fields $L^i$,
\beq
 \tilde K  = 3 -\log {\rm Im} \tau + \log \left( \tfrac 16 \cK_{ijk} L^iL^jL^k\right) 
  + \tfrac {1}{16}    \zeta(3)   \,  \,  \chi \,
 \left(
- 2 b_1 + 4 b_{2}
  \right)({\rm Im} \tau)^{3/2}\, \sqrt{ \tfrac 16 \cK_{ijk} L^iL^jL^k}  \ ,
\eeq
which has manifestly a different
functional form as compared to the zeroth-order result. Note indeed that
$-2b_1 + 4  b_{2} = 8$.

Let us now summarize the expressions for
the K\"ahler coordinates and potential
after plugging in the values of the $b$ coefficients given in \eqref{b_coefficients}.
We have
\begin{align}  
\tau_0 & = i e^{-\phi}  \ ,  \label{final1}  \\
T_i & = i\, \cK_i   \ ,  \label{final2}   \\
K & =  \phi -2 \log \left[ 
\frac{\cV}{(2\pi \alpha')^3}  - \frac 14 \,   \zeta(3) \,  e^{-\frac 32 \phi} \, \chi 
\right]  \ ,   \label{final3} \\
L_i & = \frac{v^i}{\cV} \left[
1 +  \frac 14  \,    \frac {(2\pi \alpha')^3}{ \cV} \, \zeta(3) \, e^{-\frac 32 \phi}\, \chi 
\right] \ , \label{final5}   \\
L^i \, {\rm Im}T_i & = 3 \left[
1 + \frac 14 \,    \frac {(2\pi \alpha')^3}{ \cV} \, \zeta(3) \, e^{-\frac 32 \phi}\, \chi 
 \right] \ .   \label{final6} 
\end{align}
Clearly these expressions are valid up to $\alpha'$ corrections
of higher order than the $\alpha'^3$ corrections under examination.
We have utilized the explicit expression \eqref{tildechi_expression} for $\tilde \chi$
and  we have used $\cV$ for the dimensionful volume
of the threefold, thus reinstating explicitly
all factors of $\alpha'$ for the convenience of the reader.
As we can see, the leading correction to $\tau_0$, $T_i$ is vanishing.
Let us also remark that our finding for the correction to the $\cN = 1$
K\"ahler potential is in perfect agreement with the corresponding result in 
BBHL \cite{Becker:2002nn}.

\subsection{Some comments}

Some comments about the results of the previous section are in order.
We consider the 4d, $\cN = 1$ theory arising from a Calabi-Yau
orientifold with $O7$-planes and $D7$-branes,
but we do not take into account explicitly the backreaction
of these extended objects.  For instance,
the Chern-Simons part of the effective action
for a $D7$-brane wrapping a divisor  in the Calabi-Yau threefold 
contains higher-curvature terms which are essential
in the derivation of the
contribution of $D7$-branes to the $D3$-brane tadpole cancellation condition \cite{Giddings:2001yu}.
From the point of view of the uplift to F-theory, 
$D7$-branes and $O7$-branes are encoded in the Calabi-Yau
fourfold geometry, and the metric Ansatz for dimensional reduction to
three dimensions is modified by higher-derivative corrections \cite{Grimm:2014xva,
Grimm:2014efa,Grimm:2015mua,Grimm:2013gma,Grimm:2013bha}.
These considerations suggest that  $D7$-branes and $O7$-branes
could play an important role 
in the context of higher-derivative corrections
to the bulk $\cN =1$ K\"ahler potential, which is however
beyond the scope of this  work and left for future investigation.
 
It is interesting to observe that our computation shows that the leading 
correction to the $\cN = 1$ K\"ahler coordinates $\tau_0$, $T_i$
is vanishing. As far as the 4d axio-dilaton $\tau_0$ is concerned,
this result is consistent with the expectations from F-theory, since in the context
of F-theory compactification the axio-dilaton is identified with one of the 
complex structure moduli of the elliptically fibered Calabi-Yau fourfold,
and is therefore expected not to receive corrections from the K\"ahler moduli sector
proportional to the Euler number $\chi$ of the base threefold.

\section{Conclusions} \label{conclusions_section}

In this paper we  revisited the leading $\alpha'^3$ perturbative correction
to the K\"ahler potential of a 4d, $\cN = 1$ orientifold compactification of 
type IIB superstring theory on a Calabi-Yau threefold.
We reproduced the result of Becker, Becker, Haack, and Louis (BBHL) \cite{Becker:2002nn}
by showing that a correction to the K\"ahler potential emerges,
which is proportional to the Euler characteristic of the leading order Calabi-Yau
threefold background.

 Our derivation is based on an explicit Kaluza-Klein reduction
of the type IIB bulk action from ten to four dimensions. 
The 10d two-derivative action is supplemented by the well-known  gravitational
correction of the schematic form $(t_8 t_8 + \tfrac 18\epsilon_{10} \epsilon_{10})R^4$ \cite{Grisaru:1986dk, Grisaru:1986kw, Schwarz:1982jn,
Gross:1986iv, Gross:1986mw, Sakai:1986bi,   Abe:1987ud,
Kehagias:1997cq, Kehagias:1997jg, Minasian:2015bxa},
as well as by related dilaton couplings. Since the latter are not
completely known, we parametrized them in terms of three constants
$\tilde c_1$, $\tilde c_2$, $\tilde c_3$,
see \eqref{Rreplacement}.
Supersymmetry considerations in four dimensions
allowed us to derive the linear relation \eqref{linear_relation} among these
parameters in the 10d action,
which could be checked against proposals in the literature for the
complete axio-dilaton $\alpha'^3$ sector.

The background geometry used in the dimensional reduction at two-derivative level
must be modified in order to solve the higher-dimensional equations of motion
in the presence of the $\alpha'^3$ corrections under examination.
We presented an explicit solution for this backreacted background
in terms of the non-harmonic part of the third Chern  form $c_3$
of the  zeroth order Calabi-Yau threefold. 
The $\alpha'$-corrected internal metric is still K\"ahler but no longer 
Ricci-flat. As a result, the manifold is no longer a Calabi-Yau threefold with $SU(3)$ holonomy,
but 
rather an almost Calabi-Yau threefold \cite{Joyce:2001xt} endowed with an 
 $SU(3)$ structure for which the only
non-zero torsion class is $\cW_5$ in the notation of \cite{House:2005yc}.
We also found that the entire 10d background metric
is corrected at order $\alpha'^3$ by an overall Weyl factor  proportional
to the Euler density of the zeroth order Calabi-Yau threefold. This is in analogy with the
results of \cite{Grimm:2014xva , Grimm:2014efa, Grimm:2015mua} in the context of three-dimensional M-theory vacua
in the presence of higher-derivative corrections.

The dimensional reduction is performed by allowing for small
spacetime dependent fluctuations $\delta v^i$ of the K\"ahler 
moduli of the threefold, for arbitrary $h^{1,1}$. We have showed that 
 the outcome of the dimensional reduction
can be uplifted from infinitesimal variations to a full non-linear dependence
on the moduli space. The latter is captured by simple topological
expressions proportional to the Euler characteristic of the threefold.

The resulting kinetic terms for the dilaton and the K\"ahler moduli
are consistent with 4d, $\cN = 2$ supersymmetry,
once the appropriate $\cN = 2$ field variables are identified.
The prepotential for the special K\"ahler  geometry of the
K\"ahler moduli space is found to be corrected by the expected
term proportional to the Euler characteristic of the threefold \cite{Candelas:1990qd}.

The kinetic terms in the 4d effective action can also be 
  truncated in accordance with the standard O3/O7
orientifold projection, yielding a theory with $\cN = 1$ supersymmetry.
We reproduce the 
expected BBHL correction to the  $\cN =1$ K\"ahler potential
proportional to the Euler characteristic of the classical Calabi-Yau threefold.
We also computed the leading $\alpha'$ corrections to the 
K\"ahler coordinates as a function of the threefold K\"ahler moduli,
and found that they are vanishing.

Our investigation can be extended in several directions.
It would be desirable to switch on additional type IIB fields,
for instance the axion $C_0$, 
and examine their kinetic terms to confirm the structure of the K\"ahler potential.
It would be also interesting to repeat a similar analysis for the complex
structure moduli sector, even though no analogous correction to the K\"ahler 
potential is expected. 
Finally,
it would be interesting to consider explicitly the effect of both local sources and 
bulk higher-derivative corrections on the background solution.
To pursue these directions further, however, a preliminary investigation
of the complete $\alpha'^3$-corrected type IIB bulk action would be probably 
necessary.
By a similar token, the detailed knowledge of the $\alpha'$-corrected
10d gravitino and dilatino supersymmetry variations would allow
a direct check of supersymmetry of the background in the presence of
higher-derivative corrections.

\vspace*{.5cm}
\noindent
\subsection*{Acknowledgments}

We would like to thank Ralph Blumenhagen, Thomas Grimm, Thomas Hahn, Kilian Mayer and Raffaele Savelli for interesting discussions and correspondence.  
 M.W. would like to express his thankfulness to the theoretical high-energy physics groups of the center for the fundamental laws of nature in Harvard, and the Max-Planck institute for physics in Munich, for their hospitality  during my visits.   This work was supported by the WPI program of Japan. This work was supported in part by the US National Science Foundation under award PHY-1316617.

\newpage

\begin{appendix}
\vspace{2cm} 
\noindent {\bf \LARGE Appendices}
\addcontentsline{toc}{section}{Appendices}

\section{\bf Conventions, definitions, and identities} \label{Conv_appendix}

In this work we denote 10d spacetime indices by capital Latin letters $M,N = 0,\ldots,9$,
4d spacetime indices by  $\mu,\nu = 0,1,2,3$, and the internal complex indices by $m,n,p=1,...,3$ and $\bar m, \bar n,\bar p =1, \ldots,3$.
We occasionally also make use of real internal indices $a,b = 1, \dots, 6$.
The metric signature of the ten-dimensional space  is $(-,+,\dots,+)$.
Our conventions for the totally 
anti-symmetric tensor in Lorentzian signature
 in an orthonormal frame are $\epsilon_{012...9} = \epsilon_{0123}=+1$. 
The epsilon tensor in $d$ dimensions then satisfies
\ba
\epsilon^{R_1\cdots R_p N_{1 }\ldots N_{d-p}}\epsilon_{R_1 \ldots R_p M_{1} \ldots M_{d-p}} &= (-1)^s (d-p)! p! 
\delta^{N_{1}}{}_{[M_{1}} \ldots \delta^{N_{d-p}}{}_{M_{d-p}]} \,, 
\ea
where  $s=0$ if the metric has Euclidean signature and $s=1$ for a Lorentzian metric.

We adopt the following conventions for the Christoffel symbols and Riemann tensor 
\ba
\G^R{}_{M N} & = \fr12 g^{RS} ( \pa_{M} g_{N S} + \pa_N g_{M S} - \pa_S g_{M N}  ) \, , &
R_{M N} & = R^R{}_{M R N} \, , \nonumber\\
R^{M}{}_{N R S} &= \pa_R \G^M{}_{S N}  - \pa_{S} \G^M{}_{R N} + \G^M{}_{R  T} \G^T{}_{S N} - \G^M{}_{ST} \G^T{}_{R N} \,, &
R & = R_{M N} g^{M N} \, , 
\ea
with equivalent definitions on the internal and external spaces.  
Differential $p$-forms are expanded in a basis of differential one-forms as
\beq
\Lambda = \frac{1}{p!} \Lambda_{M_1\dots M_p} dx^{M_1}\wedge \dots \we dx^{M_p} \;\; .
\eeq
The wedge product between a $p$-form $\Lambda^{(p)}$ and a $q$-form $\Lambda^{(q)}$ is given by
\beq
(\Lambda^{(p)} \we \Lambda^{(q)})_{M_1 \dots M_{p+q}} = \frac{(p+q)!}{p!q!} \Lambda^{(p)}_{[M_1 \dots M_p} \Lambda^{(q)}_{M_1 \dots M_q]} \;\; .
\eeq
Furthermore, the exterior derivative on a $p$-form  $\Lambda$ reads 
\beq
 ( d \Lambda)_{N M_1\dots M_p} = (p+1) \pa_{[N}\Lambda_{ M_1\dots M_p]} \;\;,
\eeq
while the Hodge star of   $p$-form  $\Lambda$ in $d$ real coordinates is given by
\beq
(\ast_d \Lambda)_{N_1 \dots N_{d-p}} = \frac{1}{p!} \Lambda^{M_1 \dots M_p}\epsilon_{M_1 \dots M_p N_1\dots N_{d-p}} \;\; .
\eeq
Moreover, the identity
\beq\label{idwestar}
 \Lambda^{(1)} \we \ast \Lambda^{(2)} = \frac{1}{p!}\Lambda^{(1)}_{M_1\dots M_p} \Lambda^{(2)}{}^{M_1\dots M_p} \ast 1 \;\;
\eeq
  holds  for two arbitrary $p$-forms $\Lambda^{(1)}$ and $\Lambda^{(2)}$.

Let us specify in more detail our conventions regarding complex coordinates
in the internal space.
For a 
complex   manifold $M$ with complex dimension $n$
the complex coordinates $z^1 , \dots, z^n$ and 
the underlying real coordinates $y^1, \dots , y^{2n}$ are related by
\begin{equation}
( z^1,...,z^n ) = \left(  \tfrac{1}{\sqrt{2}}(y^1 + i y^2), \dots ,  \tfrac{1}{\sqrt{2}}(y^{2n-1} + i y^{2n}) \right) \,.
\end{equation}
Using these conventions one finds
\begin{equation}
\sqrt{g}  \, dy^1 \wedge ... \wedge dy^{2n} = \sqrt{g} (-1)^{\frac{(n-1)n}{2}} i^n  dz^1\wedge...\wedge dz^n 
\wedge d\bar z^1 \wedge...\wedge d\bar z^n = \frac{1}{n!} J^n \,,
\end{equation}
with $g$ the determinant of the metric $g_{ab}$ in real coordinates and  $\sqrt{\det g_{ab}} = \det g_{m \bar n} $. The K\"{a}hler form is given by
\begin{equation}
\label{eq:Kform}
J = i g_{m\bar{n} } dz^m \wedge d\bar z^{\bar{n} } \, .
\end{equation}
Let $\omega_{p,q}$ be a $(p,q)$-form, then its Hodge dual is the $(n-q,n-p)$ form
 \begin{multline} \label{eq:pgform}
\ast \omega_{p,q}  = \frac{ (-1)^{\frac{n(n-1) }{2}  } \, i^n \, (-1)^{pn}}  {p!q!(n-p)!(n-q)!}   
\;
 \omega_{m_1 \dots m_p \bar{n} _1 \dots \bar{n} _q} 
\epsilon^{m_1 \dots m_p}_{\phantom{m_1 \dots m_p} \bar r_1 \dots \bar r_{n-p}}  \epsilon^{\bar{n} _1 \dots \bar{n} _q}_{\phantom{\bar \beta_1 \dots \bar{n} _q}  s_1 \dots  s_{n-q}} 
\, \times \nn \\
 \times \,
dz^{ s_1}\wedge \dots \wedge dz^{ s_{n-q}} \wedge d \bar z^{\bar r_1} \wedge \dots \wedge d \bar z^{\bar  r^{n-p}}.
\end{multline}
Finally, let us record our conventions regarding Chern forms.
To begin with, 
 we define the curvature two-form for Hermitian manifolds to be
  \begin{equation}\label{curvtwo}
 {\cR^m}_n  =  {{R^m}_n }_{ r \bar s} \, dz^ r \wedge d\bar{z}^\bar{s}\;\;,
  \end{equation}
and we set
 \bea \label{defR3}
 \Tr{\cR}\;\;&  =& {{R^ m }_ m }_{ r_1 \bar{s}_2 }  \, dz^ {r_1} \wedge d\bar{z}^{\bar{s_1}} \;,\nonumber \\
 \Tr{\cR^2} &= & {{R^{ m }}_{n }}_{ r_1 \bar{s_1}}  \,  {{R^{n }}_{ m }}_{ r_2 \bar{s}_2} \, 
  dz^{ r_1} \wedge d\bar{z}^{\bar{s}_1}
 \wedge dz^{ r_2} \wedge d\bar{z}^{\bar{s}_2} \;,\nonumber  \\
 \Tr{\cR^3} &=& {{R^{ m }}_{n }}_{ r_1 \bar{s_1}} \, 
  R^{n }{}_{p  r_2 \bar{s}_2} \,
 {{R^{p}}_{ m }}_{ r_3 \bar{s}_3}    \, 
 \wedge dz^{ r_1} \wedge d\bar{z}^{\bar{s}_1}
 \wedge dz^{ r_2} \wedge d\bar{z}^{\bar{s}_2}
  \wedge dz^{ r_3} \wedge d\bar{z}^{\bar{s}_3} \; .
 \eea
 The Chern forms can then  be expressed in terms of the curvature two-form as
\begin{align}  \label{Chernclasses}
 c_0 &= 1 \nonumber \;, \\
 c_1 &= \frac{1}{2\pi} i \Tr{ \mathcal{R}} \nonumber\;, \\
 c_2 &= \frac{1}{(2\pi)^2} \frac{1}{2}\left( \Tr{\cR^2} -  (\Tr{\cR})^2 \right)\;, \\
 c_3 &=   \frac{1}{3}c_1c_2 + \frac{1}{(2\pi)^2} \frac{1}{3} c_1 \wedge \Tr \cR^2 - 
 \frac{1}{(2\pi)^3}\frac{i}{3} \Tr \cR^3 
   \;,\nonumber \\
 c_4 &= \frac{1}{24} \left( c_1^4 - \frac{1}{(2\pi)^2} 6c_1^2 \Tr\cR^2 
 - \frac{1}{(2\pi)^3} 8i c_1 \Tr\cR^3\right) 
 + \frac{1}{(2\pi)^4} \frac{1}{8}((\Tr\cR^2)^2 - 2 \Tr\cR^4)   \,. \nonumber 
\end{align}
The Chern forms of an $n$-dimensional Calabi-Yau manifold $Y_n$ reduce to
\beq\label{chern34}
c_3 (Y_{n \geq 3}) =  - \frac{1}{(2\pi)^3} \frac{i}{3}  \Tr{\cR^3} \;\; \text{and} \;\;
 c_4 (Y_{n \geq 4}) =\frac{1}{(2\pi)^4} \frac{1}{8}((\Tr\cR^2)^2 - 2 \Tr\cR^4)\;.
\eeq

\section{Higher-derivative dilaton terms}\label{app action}
In this appendix we record the expression of the quantities
$f_1$, $f_2^{MN}$, $f_3$ introduced in \eqref{R4actiontot}.
We have
\ba
f_1 = & 192   R_{ M}{}^{R}{}_{O}{}^{S}R^{ M N O P}R_{N S P R} -48 R_{ M N}{}^{R S}R^{ M N O P}R_{O P R S} & \nonumber  \\ 
&+ 576   R^{ MN}{}_{ M}{}^{O}R_{N}{}^{P R S}R_{O P R S}  +384 R^{ M N}{}_{ M}{}^{O}R_{N}{}^{P}{}_{P}{}^{R}R_{O}{}^{S}{}_{R S}& \nonumber  \\ 
&-72  R^{ M N}{}_{ M N}R_{O P R S}R^{O P R S} -576   R^{ M N}{}_{ M}{}^{O}R_{N}{}^{P}{}_{O}{}^{R}R_{P}{}^{S}{}_{RS}& \nonumber  \\ 
&+ 288  R^{ M N}{}_{ M N}R^{O P}{}_{O}{}^{R}R_{P}{}^{S}{}_{R S} -24  R^{ M N}{}_{ M N}R^{O P}{}_{O P}R^{R S}{}_{R S} \  ,\\[3mm]
f_{2}^{MN} =  g^{MN}( &1 92 R_{R}{}^{Q}{}_{O}{}^{V}R^{R S O P}R_{S V P 
Q}-48 R_{R S}{}^{Q V}R^{R S O P}R_{O P Q V}& \nonumber  \\ 
&+576 R^{R S}{}_{R}{}^{O}R_{S}{}^{P Q V}R_{O P Q V}
+384  R^{R S}{}_{R}{}^{O}R_{S}{}^{P}{}_{P}{}^{Q}R_{O}{}^{V}{}_{Q 
V}& \nonumber  \\ 
&-72  R^{R S}{}_{R S}R_{O P Q V}R^{O P Q V}
-576  R^{R S}{}_{R}{}^{O}R_{S}{}^{P}{}_{O}{}^{Q}R_{P}{}^{V}{}_{Q V}& \nonumber  \\ 
&+288  R^{R S}{}_{R S}R^{O P}{}_{O}{}^{Q}R_{P}{}^{V}{}_{Q V}
-24   R^{R S}{}_{R S}R^{O P}{}_{O P}R^{Q V}{}_{Q V}) + \dots  
= g^{MN} \, f_1 + \dots&
\ea
where the ellipsis denote terms where the free indices $M,N$ are on the Riemann tensors and thus will not contribute to our discussion. Furthermore,
\ba
f_3=  & 768   R_{M O}{}^{R S}R^{M N O P}R_{N R P S}
+384  R_{M}{}^{R}{}_{O}{}^{S}R^{M N O P}R_{N S P R} & \nonumber \\ &
 -96  R_{M N}{}^{R S}R^{M N O P}R_{O P R S}
+1536 R^{M N}{}_{M}{}^{O}R_{N}{}^{P R S}R_{O P R S}& \nonumber \\ &
-384   R_{M N}{}^{R S}R^{M N O P}R_{O R P S}
+768  R^{M N}{}_{M}{}^{O}R_{N}{}^{P R S}R_{O R P S}& \nonumber \\ &
   +1536   R^{M N}{}_{M}{}^{O}R_{N}{}^{P}{}_{P}{}^{R}R_{O}{}^{S}{}_{R S}
   -240  R^{M N}{}_{M N}R_{O P R S}R^{O P R S}& \nonumber \\ &
   -1920   R^{MN}{}_{M}{}^{O}R_{N}{}^{P}{}_{O}{}^{R}R_{P}{}^{S}{}_{R S}
+1152   R^{M N}{}_{M N}R^{O P}{}_{O}{}^{R}R_{P}{}^{S}{}_{R S}& \nonumber \\ &
-96  R^{M N}{}_{M N}R^{O P}{}_{O P}R^{R S}{}_{R S}   \ .&
\ea

\end{appendix}

\bibliographystyle{JHEP}

\bibliography{references}

\end{document}